\documentclass[preprint,12pt]{elsarticle}

\usepackage{graphicx, subfigure}

\usepackage{amssymb}

\journal{\tt{arXiv.org}}

\begin{document}

\begin{frontmatter}


\title{Natural vs.~blocked ventilation in naturally ventilated buildings -- the effect of finite and decoupled sources}



\author{Vincent Mayoraz$^{1,2}$ and M.\,R.~Flynn$^2$}

\address{$^{1}$ Section de G{\' e}nie M{\' e}canique, Ecole polytechnique f{\' e}d{\' e}rale de Lausanne, Lausanne, Switzerland \\
$^{2}$Dept.~of Mechanical Engineering, University of Alberta, Edmonton, AB, Canada}

\begin{abstract}
Naturally ventilated buildings harness freely-available resources such as internal buoyancy gains and wind forcing in achieving comfortable interior conditions. Although these resources are free, they are time-variable and can be difficult to control. As a result of this the nonlinear interplay between sometimes competing resources may lead to sub-optimal ventilation states. This problem has been explored by a number of previous researches e.g.~Flynn \& Caufield ({\it Building and Environment}, {\bf 44}, 216--226, 2009) who, in studying these ventilations states, demonstrated complicated transitions characterized by hysteresis even for the simple case of a one-zone building. The objectives of this research are to extend the previous (theoretical) analysis by conducting complementary numerical experiments using sophisticated algorithms capable of describing turbulent, buoyancy driven flow. A further objective of this study is to specifically “decouple” the source, which was previously assumed to supply both heat and mass to the interior space. Rather, the flow dynamics in a case where heat and mass are supplied independently are examined.
\end{abstract}

\begin{keyword}
Fluid mechanics \sep Natural ventilation \sep Numerical simulations 


\end{keyword}

\end{frontmatter}



\setlength\parindent{0pt}

\section{Introduction}  

``Naturally-ventilated'' buildings are a special class of net-zero energy buildings that seek to reduce the energy expense by harnessing freely-available resources (solar radiation, wind forcing, internal buoyancy generation) in forcing air into and out of the built environment. Although the previously mentioned resources are free, they are, with the possible exception of patterns of building occupation and the switching on and off of electrical and related equipment, unpredictable. Strictly regulating interior conditions to ensure that they fall within a range deemed comfortable is therefore no trivial task. In this sense, it is important to understand the flow dynamics within the buildings, in order that appropriate control schemes may be devised. Numerous studies have tried to propose practical equations, giving the architects/designers simple rules of thumb that nonetheless capture the essential physics governing architectural flows \cite{kayeflynn2012}.\\

In parallel, researchers have used more sophisticated numerical simulations to predict the flow behavior of natural ventilation \cite{Kaye2009}. Note, however, that numerically studies have generally been limited to certain specific scenarios or flow states and not to the whole range of air flows that can take place in naturally ventilated buildings. Therefore, one of the major objectives of the present research is to examine the details of the unexpected flow behavior when the thermal source is ``non-ideal,'' i.e. it supplies both heat and mass to the interior space. This particular research interest is one piece of a much larger puzzle aimed at understanding how the principles of natural ventilation, originally developed for application in temperate climates, can be profitably extended to a broader range of locales. In so doing, the overarching objective is to reduce so much as possible the need for expensive and energy-intensive mechanical systems, which have for too long characterized building design. Natural ventilation offers an appealing alternative to traditional practice insofar as the waste heat produced during summer months can be used to remove stale air to be replaced by fresh outside air. By contrast, in winter that heat, no longer unwanted, can be used to generate controlled air flows helping to maintain comfortable temperatures inside.\\

Finally this research will try to show that it is possible to perform good quality studies of complex flow phenomenon inside naturally ventilated buildings using commercial CFD softwares on standard workstations. In the long term, this could open the door to a popularization of naturally ventilated buildings since the design of such buildings was until not so far ago too complex to be applied to homes and commercial buildings, this due to time limitations and monetary constraints.


\section{The analytical model}

\subsection{Description of the problem} 
Due to the complexity of the problem even for simplified geometries, natural ventilation needs to be modeled. In any case these models can roughly be divided in five different types, taking into account -or not - different parameters influencing the flow. These parameters and the associated models are summarized in the Table \ref{tab:general} and are detailed in the following paragraphs.

\begin{table}[h]
\centering
\begin{tabular}{r c c c l}

	Case & $P$ & $Q_s$ & W & Experimental validation \\
    
    \hline
	1 & $>$ 0 & = 0 & = 0 & \citet{Kaye2004} \\
	2 & $>$ 0 & = 0 & $>$ 0 & \citet{Flynn2009} \\
	3 & = 0 & $>$ 0 & = 0 & \citet{Woods2003} \\
	4 & = 0 & $>$ 0 & $>$ 0 & - \\
    5 & $>$ 0 & $>$ 0 & $>$ 0 & - \\
    \hline
\end{tabular}
\caption{Different models and their parameters. $P$ is the power of the heat source, $Q_s$ is the source volume flux and W is the speed of the wind}
\label{tab:general}
\end{table}

\noindent The objective of this section is not only to highlight and show the main relevant studies carried out in the field but also to define the problem and the associated quantities that will be used further together. It will give an idea of the limitations of the mentioned models. The following section emphasizes the main results that will further be compared with the CFD simulations and is by no means a comprehensive list of the existing prediction models.\\

\subsection{The emptying filling box model} \label{sec:ideal-no-wind-theo}
\begin{figure}[ht]
\centering
  \includegraphics[width=\textwidth]{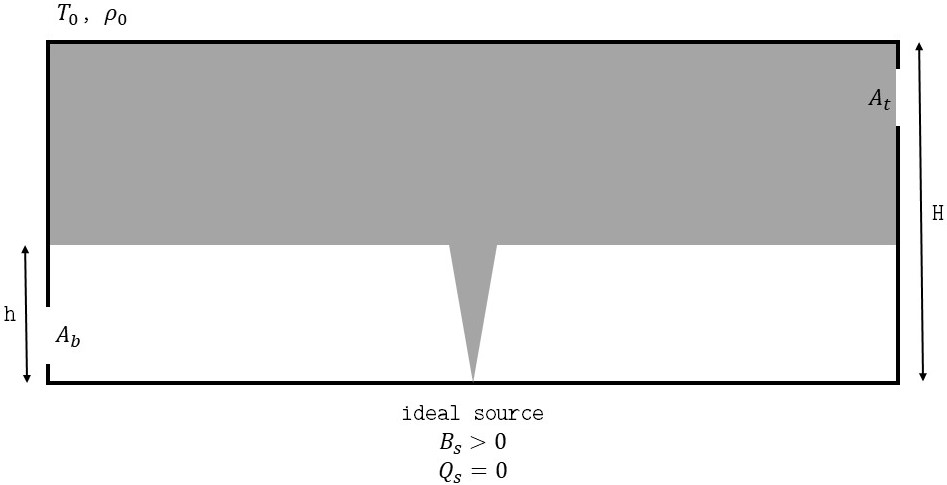}
  \caption{Illustration of the considered idealized geometry and its relevant parameters.}
  \label{fig:one_box_building}   
\end{figure}

The emptying filling box model considers an empty box with a floor surface area $S$, height $H$ and bottom and top openings area $A_b$ respectively $A_t$. A plume rises from a source to the top of the box due to thermal differences and spreads out laterally to form a density interface between the plume outflow and the ambient fluid. \citet{Kaye2004} have been studying this flow using the ``filling box'' model developed by \citet{Baines1969} and \citet{Linden1990}. The source $P$ is considered ideal since only heat and no mass is added to the system. The outside temperature and density are fixed to $T_0$ and $\rho_0$ as shown in Fig.~\ref{fig:one_box_building}.
 
\noindent Different properties of the inner flow can be predicted by theory, in function of the initial parameters and the geometry of the box. The ``filling time,'' i.e. the time for the ventilated box with a single source to be filled with buoyant fluid, can be expressed as

\begin{equation}
t_f = \frac{S}{\lambda B^{1/3} H^{2/3} }
\end{equation}

\noindent On the other hand, the ``draining time,'' i.e. the time taken to flush the box of the aforementioned fluid can also be predicted:

\begin{equation}
t_d = \frac{ \lambda^{1/2} H^{4/3} S }{ A^* B^{1/3} }
\end{equation}

\noindent Both of these times depend on the constant $\lambda$ which is widely used in the modeling of plumes and is equal to

\begin{equation}
\lambda = \frac{6 \alpha}{5} \left( \frac{9 \alpha}{10} \right)^{1/3} \pi^{2/3}
\end{equation}

\noindent where the coefficient $\alpha$ is the plume entrainment coefficient. It depends on the geometry of the plume and in this case (asymmetric plume with an assumed top-hat distribution for vertical velocity and density) is about $\sim$ 0.114. $B$ represents the plume buoyancy flux created by the temperature difference. It is expressed in function of gravity, the heat source power $P$ and the characteristic properties of the fluid:

\begin{equation} \label{equ:buoyancy1}
B = \frac{ Pg }{ c_p \rho_0 T_0 }
\end{equation}

\noindent $A^*$ is a non-dimensional parameter called the effective area defined to facilitate the comparison for different sizes and geometries of openings. It is a function of $A_t$ and $A_b$, and is expressed as

\begin{equation}
A^* = \frac{ \left( 2 c_t c_b A_t A_b \right)^2 }{ \sqrt{2c_t^2 A_t^2 + 2 c_b^2 A_b^2} }
\end{equation}

\noindent where $c_t$ and $c_b$ denote the discharge loss coefficients associated with the flow through the bottom and top openings. \citet{Hunt2000} showed that these coefficients exhibited density dependence but within the framework of this project they will be considered constant and equal to 0.6 for simplicity and since their impact on the final predictions is low.\\ 

\begin{figure}[ht]
\centering
  \includegraphics[width=0.8\textwidth]{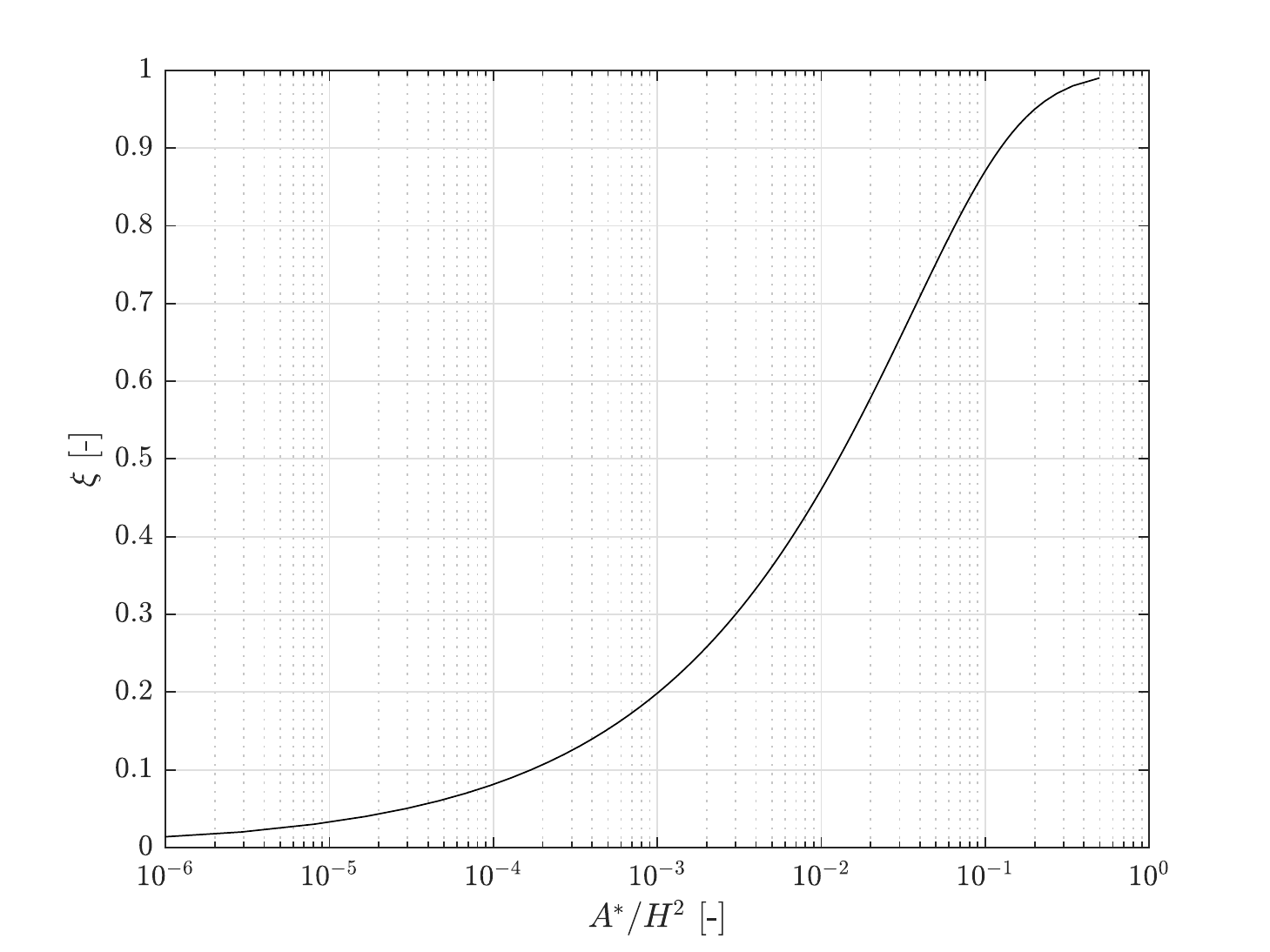}
  \caption{Variation of the non-dimensional interface height with respect to the time scale ratio with regard to the effective area.}
  \label{fig:xi_vs_mu_and_AH2}   
\end{figure}

Using the filling box model again, \citet{Linden1990} proved that in the case of a single buoyant point source the steady state interface height could be expressed in function of the normalized effective area and the coefficient $\lambda$ only:
\begin{equation} \label{equ:no_wind}
\frac{A^*}{H^2} = \lambda^{3/2} \cdot \frac{t_f}{t_d} = \frac{\lambda^{3/2} \xi^{5/2}}{\sqrt{1-\xi}}
\end{equation}
where the interface height is non-dimensionalized to ease comparison with different geometries ($\xi = h/H$). Fig.~\ref{fig:xi_vs_mu_and_AH2} represents the variation of the steady state interface height with respect to the non-dimensionalized area, which corresponds to the equ.~(\ref{equ:no_wind}). $A^*$ is ``symmetric'' in $A_t$ and $A_b$, thus the height of the interface can be modified with a change of $A_t$ or $A_b$. Small effective areas will induce lower interface heights since the size of the openings will not be large enough to allow the escape of the buoyant layer. On the other hand, very large effective areas will induce a density interface close to the roof, corresponding to a case where almost all the heated air flows out of the room.

\noindent The transient behavior of this flow has also been studied by \citet{Kaye2004} and \citet{Moradi2017}. They showed that the evolution of the interface height could be expressed by the following system of non-dimensional differential equations: 

\begin{equation}
\frac{d \xi}{d \tau} = \frac{1}{\sqrt{\mu}} \sqrt{\eta (1-\xi)} - \sqrt{\mu} \xi^{5/3}
\end{equation}
\begin{equation}
\frac{d \eta}{d \tau} = \sqrt{\mu} \ \frac{1 - \xi^{ \frac{5 \delta}{3} }}{1 - \xi}
\end{equation}

\noindent where $\tau$ represents the non-dimensional time, $\mu$ is the ratio of the draining and filling times ($\mu = t_d/t_f$) and $\eta$ represents the dimensionless average reduced gravity of the buoyant layer:

\begin{equation}
\eta = g' \frac{\lambda H^{5/3}}{B^{2/3}}
\end{equation} 

Here, the reduced gravity is defined by

\begin{equation}
g' = g \left( \frac{T_{up} - T_0}{T_0} \right) = \frac{1}{\lambda} \left[ \frac{B^2}{(\xi H)^5} \right]^{1/3}
\end{equation}

\noindent where the first equality corresponds to the general expression of the reduced gravity and the second one to the equivalent definition for natural ventilation, with regards to the interface height. 

\begin{figure}[ht]
\centering
  \includegraphics[width=0.8\textwidth]{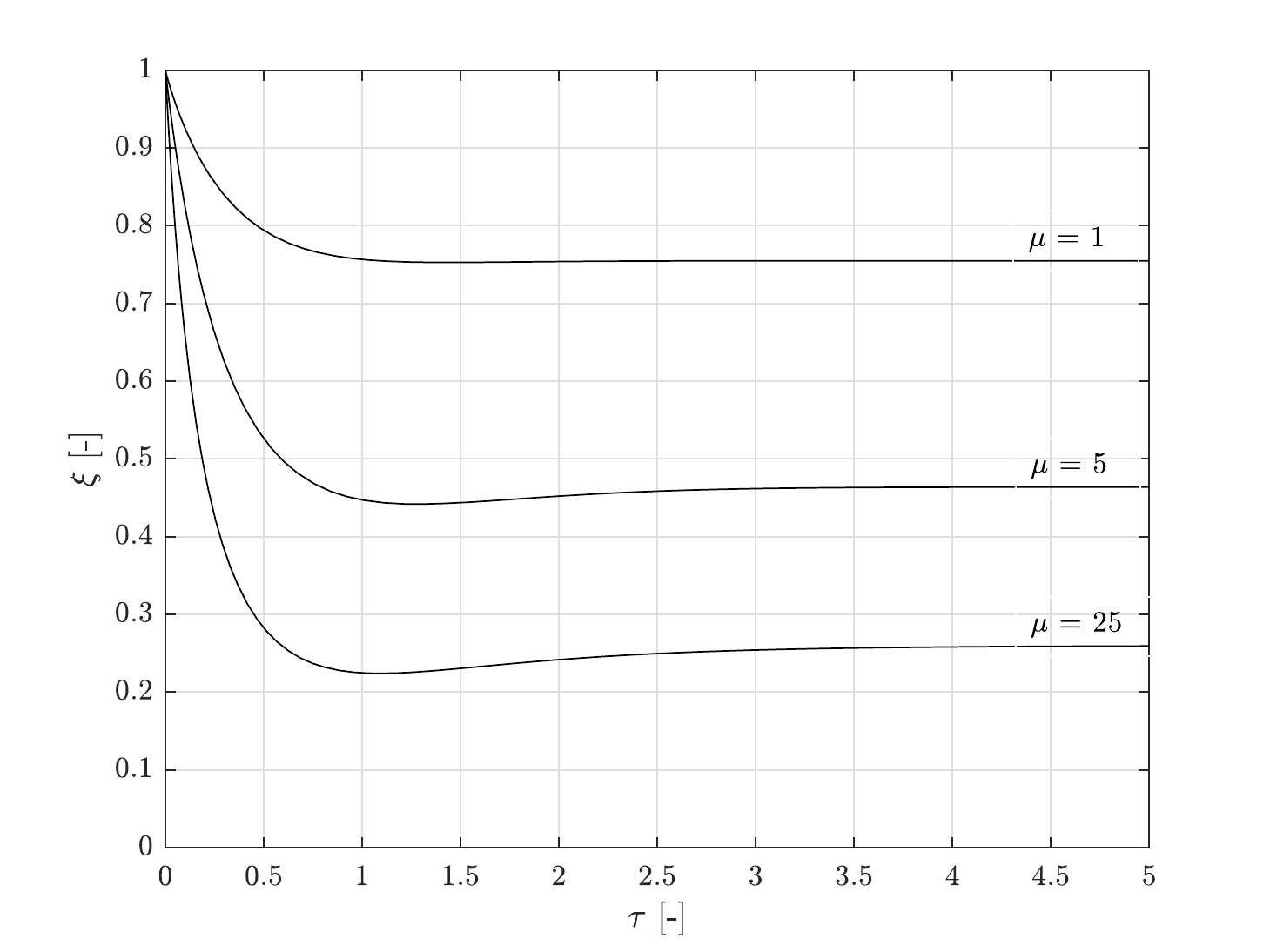}
  \caption{Evolution of the non-dimensional interface height with respect to the non-dimensional time for different values of  $\mu$.}
  \label{fig:xi_vs_tau}   
\end{figure}

\noindent Fig.~\ref{fig:xi_vs_tau} represents the evolution of the non-dimensional interface height $\xi$ with respect to the non-dimensional time $\tau$ for different values of  $\mu$. It can be seen that the interface overshoots before reaching its steady state value. The analytical model was compared with experimental data by \citet{Kaye2004} from a small-scale model using salty water to reproduce the thermal plume. The theoretical model showed good agreement with the small-scale experimental model.

\subsection{Influence of the external atmosphere} \label{sec:external-atm-theo}
\begin{figure}[ht]
\centering
  \includegraphics[width=\textwidth]{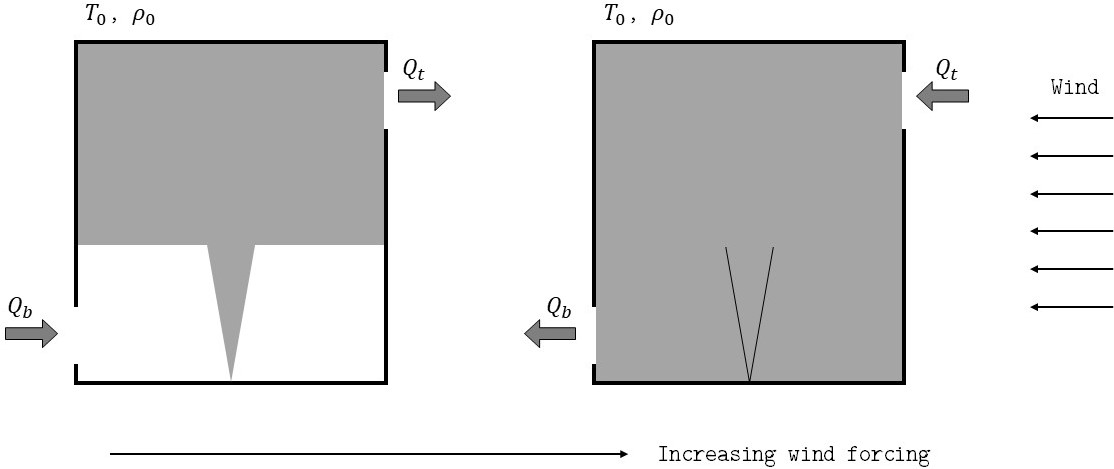}
  \caption{Schematic illustration of observable flow regimes in the presence of an adverse wind assuming a zero volume flux source of infinite temperature. In this circumstance, naturally ventilated (left) and well-mixed (right) states are possible.}
  \label{fig:wind_forcing}   
\end{figure}

To improve the realism of the previous model, the influence of the external atmosphere on the inner flows is added. The governing features of the previously mentioned flows have been studied by \citet{Linden1999}. It differentiates two main regimes that can be encountered in natural ventilation: mixing ventilation in which the interior is at an approximately uniform temperature, and displacement ventilation where there is strong internal stratification. In this sense, \citet{Flynn2009} have been exploring the role of wind, exerting different static pressures on the leeward and windward facades of the building. To do so, the idealized geometry presented in the Fig.~\ref{fig:wind_forcing} is considered. The presence of an adverse wind induces a pressure drop $\Delta p$ between the top and the bottom of the room which needs to be taken into account in the computations of the stratification layer.

\noindent  Thereby, in the case of a sufficient thermal forcing the influence of an external pressure (produced by wind) can be added to equ.~(\ref{equ:no_wind}) which becomes

\begin{equation} \label{equ:wind}
\frac{A^*}{H^2} = \frac{ \lambda^{3/2} \xi^{5/3} }{ \sqrt{ \frac{ 1-\xi }{ \xi^{5/3} } - \lambda \frac{ \Delta p}{\rho_0} \left( \frac{H}{B} \right)^{2/3}} }
\end{equation}

\begin{figure}[ht]
\centering
  \includegraphics[width=0.8\textwidth]{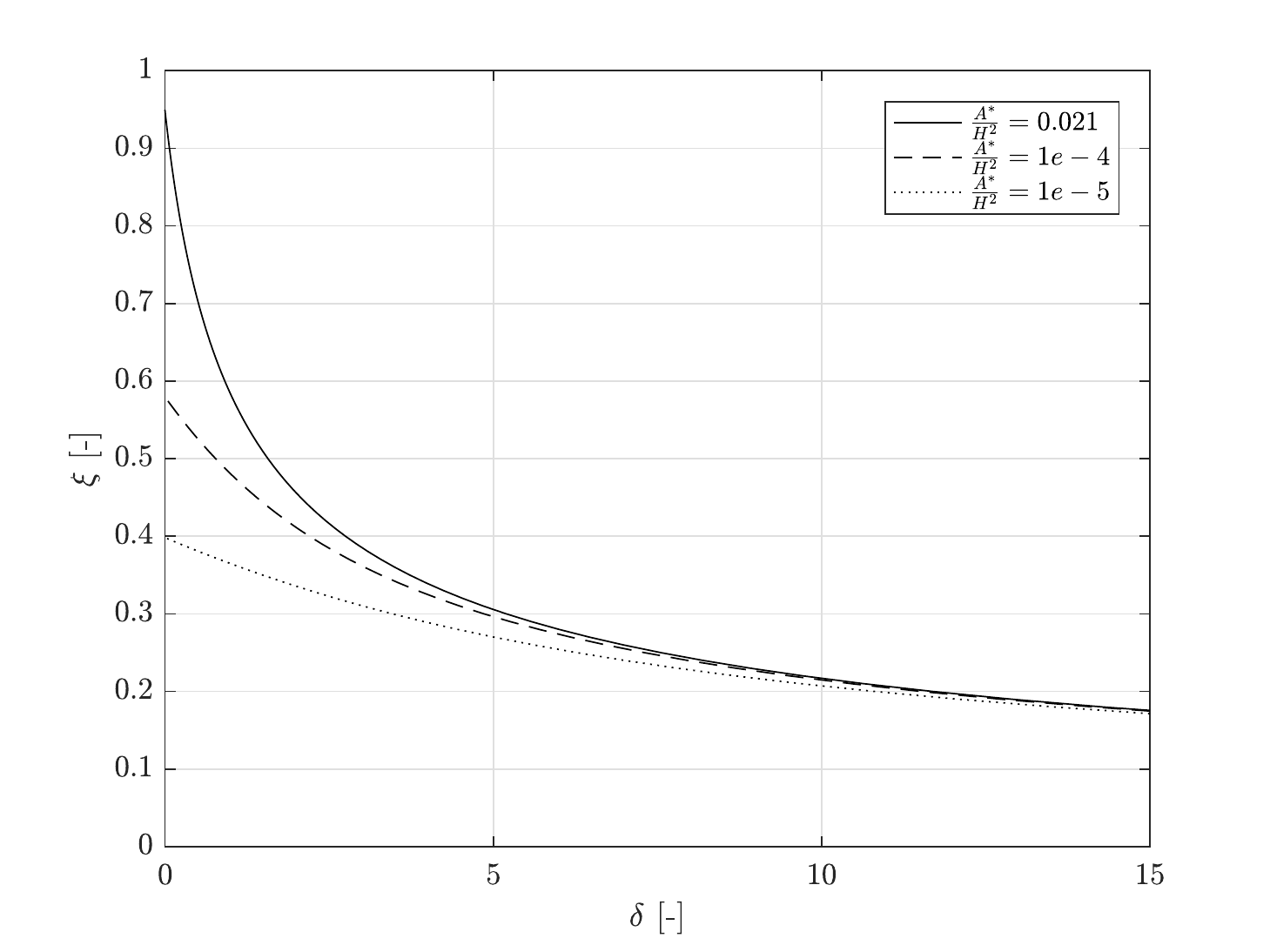}
  \caption{Variation of the non-dimensional interface height with respect to the normalized Froude number $\delta$ (equ.~\ref{equ:wind}).}
  \label{fig:xi_vs_delta}   
\end{figure}

\noindent The effect of the pressure difference is taken into account in the equation and weighted by the strength of the buoyancy flux $B$. Intuitively higher adverse wind forces will induce higher pressure differences and will tend to lower the height of the interface. On the other hand a strong buoyancy flux $B$ will compensate the effect of the pressure drop and help to raise the interface height. New variables need to be introduced in order to describe this more complex flow. In the case of a single building the Froude number $Fr$ is equal to

\begin{equation} \label{equ:froude}
Fr = \sqrt{ \frac{\Delta p}{\rho_0} \left( \frac{H}{B} \right)^{2/3} }
\end{equation}

\noindent and can be recognized as a term of equ.~(\ref{equ:wind}). Moreover,a non-dimensional number is defined: $\delta = \lambda Fr^2$ which will be used later to simplify the mathematical relations between the different parameters describing the flow. Fig.~\ref{fig:xi_vs_delta} represents the non-dimensional interface height with respect to that non-dimensional parameter $\delta$ (equ.~\ref{equ:wind}). It can be seen that the height of the interface is lowered by the action of the wind inducing a higher pressure on the inlet of the building. The influence of the effective area is diminished in the case of high values of $\delta$ which could be expected since the wind becomes the main driving parameter of the flow.

\noindent Moreover, the wind-driven and buoyancy-driven volume fluxes can be expressed as

\begin{equation}
Q_B = A^* (g' H)^{1/2} \qquad and \qquad Q_w = A^* Fr \left( \frac{B}{H} \right)^{1/3}
\end{equation}

\noindent Thereafter, the ventilation rate for the building can be computed in function of the parameters defined above, for the stratified state, respectively mixed state:

\begin{equation} \label{equ:fluxes}
Q = A^* \sqrt{g' H (1 - \xi) - \frac{\Delta p}{g' \rho_0}}, Q_w < Q_B -\sqrt{Q_w^2 - Q_B^2},  Q_w > Q_B 
\end{equation}

\noindent The flux is considered positive if the buoyancy forces are dominant, i.e. the flow will go out through the upper (top) opening and negative when the wind dominates, where the air will escape through the lower opening. \citet{Hunt2004} proved that the flow state could be predicted with the parameters defined before with a simple polynomial equation:

\begin{equation} \label{equ:states1}
\left( \frac{Q}{Q_W} \right)^3 - \frac{Q}{Q_W} - \frac{(H^2/A^*)}{Fr^3} = 0
\end{equation}

\begin{figure}[ht]
\centering
  \includegraphics[width=0.8\textwidth]{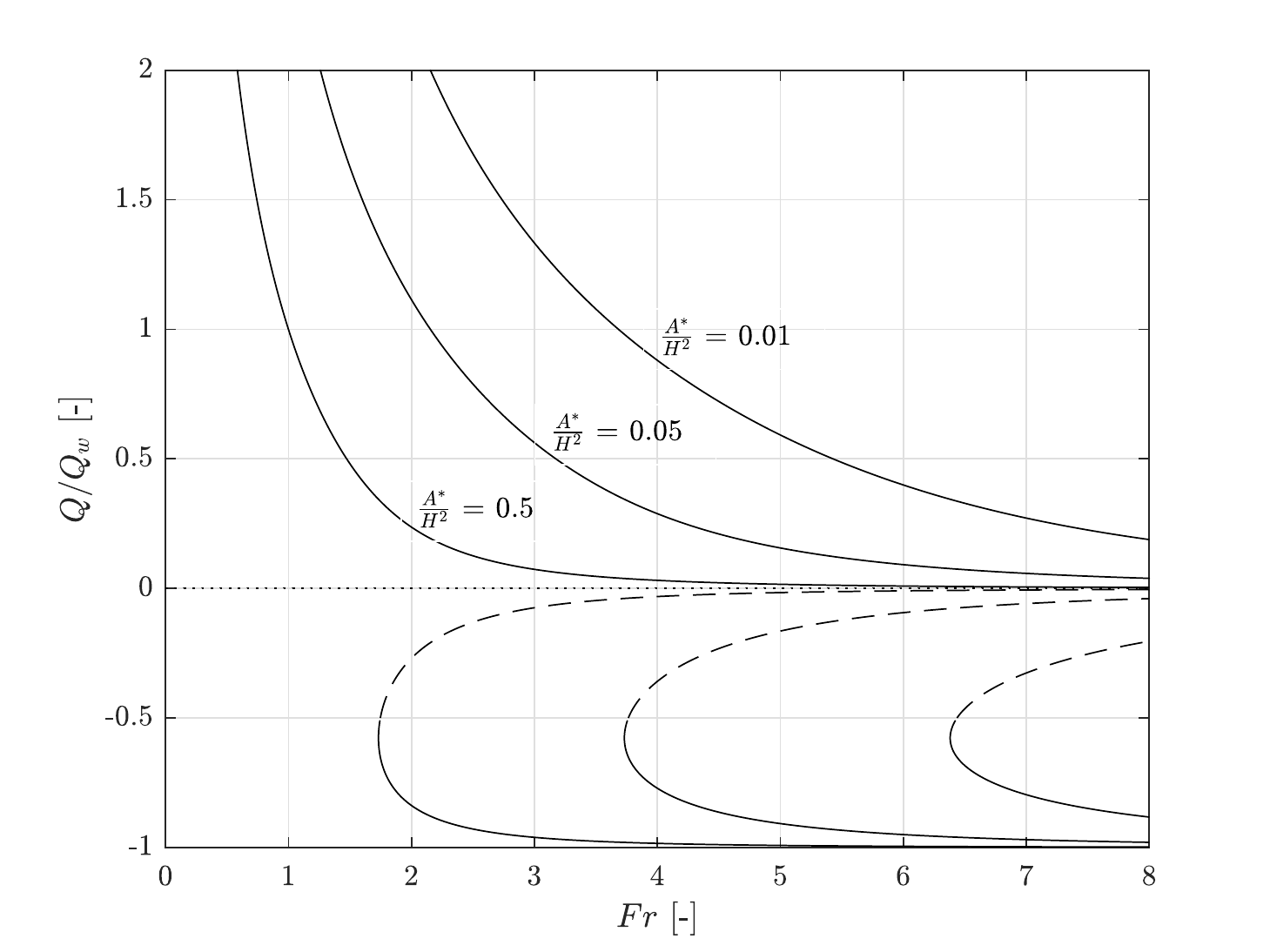}
  \caption{Non-dimensionalized total volume flow with respect to the Froude number (equ.~\ref{equ:states1}).}
  \label{fig:QQw_vs_Fr}   
\end{figure}

\noindent As shown in Fig.\ref{fig:QQw_vs_Fr} multiple solutions exist in general. Two of these solutions correspond to the well mixed state (only one of which is stable) and the other corresponds to the naturally ventilated state.\\ 

To explain the transition from mixing to stratified, two cases can be considered. On one hand a case with a constant wind and no heat source ($B$ = 0). This case corresponds to a stable mixing flow with infinite Froude number where the normalized flow rate $Q/Q_W$ = -1. When $B$ is increased from zero, $Fr$ decreases (equ.~\ref{equ:froude}), and the steady state solution falls on the lower (stable) part of the solution. The buoyancy driven flux $Q_B$ increases in magnitude with $B$ and hence $|Q/Q_w|$ decreases. At the critical Froude number corresponding to the left end of the mixed solution, the wind will be unable the maintain the flow in the room for any further increase in $B$, which will result in a transition from the mixed to the stratified regime.\\

On the other hand, if the room contains a non-zero heat source ($B$>0) and the wind speed is initially zero, the system will be located on the upper part of the solution, where $Q/Q_w$ is equal to infinity. Supposing now that the wind is progressively increased, $Fr$ will increase as $Q_w$ so that $Q/Q_w$ decreases and buoyancy accumulates in the upper part of the building. This behavior goes on until the wind is too strong and the flow, unable to maintain the stratified regime, moves to the unstable part of the equation. As explained above, since this state is unstable, the flow will move towards one or the other solution. Physically this corresponds to a hysteresical behavior of the flow at low values of $Q/Q_w$.

\subsection{Blocked ventilation} \label{sec:non-ideal-no-wind}
The critical state depicted in Fig.\ref{fig:QQw_vs_Fr}, between the well mixed and stratified state, where the flow will become hysteresical can be modeled more finely if a non-zero mass flux inlet is added at the heat source. This will result in a state where flow escapes from both top and bottom openings, whereas still respecting the mass conservation for the system. This type of regime is called ``blocked ventilation.'' The buoyancy flux defined in equ.~(\ref{equ:buoyancy1}) for a heat power source $P$ becomes for a non-zero source mass flux

\begin{equation}
B_s = g Q_s \left( \frac{T_{in}-T_0}{T_0} \right)
\end{equation}

\noindent where $Q_s$ is the mass flux in [m$^3$/s] and $T_{in}$ is the temperature of the air injected. In order to simplify the further mathematical relations, the source mass flux is non-dimensionalized as follows:

\begin{equation}
\zeta_s = \frac{1}{H} \left( \frac{Q_s}{\lambda B_s^{1/3}} \right)^{3/5}
\end{equation}

\begin{figure}[ht]
\centering
  \includegraphics[width=.8\textwidth]{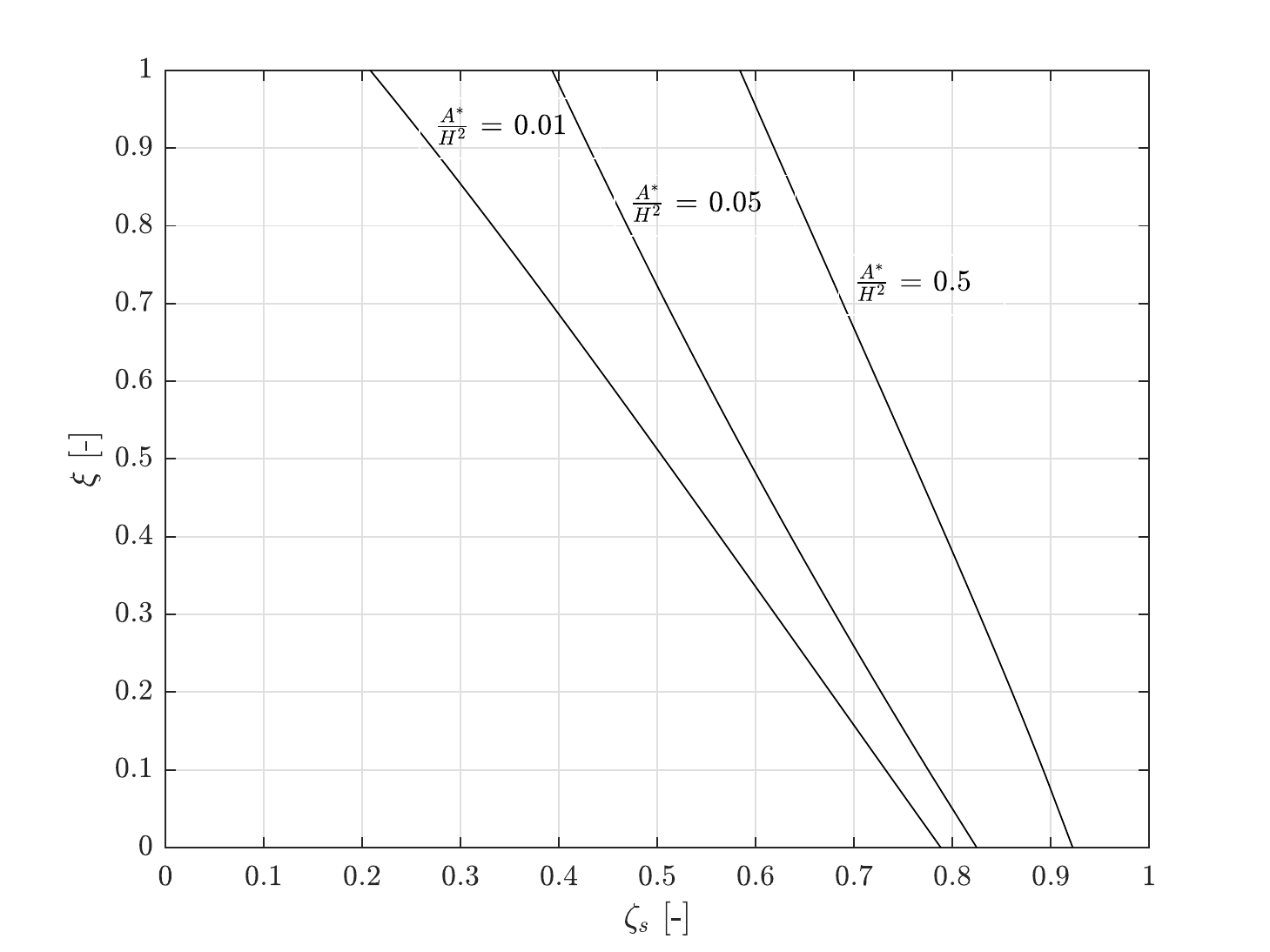}
  \caption{Variation of the non-dimensional interface height with respect to the non-dimensional source mass flux.}
  \label{fig:xi_vs_qs_woods}   
\end{figure}

\noindent Blocked ventilation and the effect of the non-ideal heat source were studied by \citet{Woods2003} among others. They carried out both analytical and experimental campaigns ending up with a validated model for negative plumes initiated by source mass fluxes. They proposed an equation linking the density interface height with the properties of the flow and the inlet parameters. Using the Boussinesq approximation, where positive and negative plumes are equivalent due to the symmetry of the driving forces, it is possible to show that the model studied by \citet{Woods2003} can be applied in the case studied here and re-written with the parameters defined above in a non-dimensional form:

\begin{equation} \label{equ:woods_modified}
\frac{A^*}{H^2} =    \frac{ \lambda^{3/2} \left[ (\xi+\zeta_s)^{10/3} - \frac{A_T^2 \left( 2 \zeta_s^{5/3} (\xi+\zeta_s)^{5/3} -\zeta_s^{10/3} \right) }{A_T^2 + A_B^2}  \right] ^{1/2}}  {\sqrt{ \frac{ 1-\xi }{ (\xi+\zeta_s)^{5/3} } } }
\end{equation}

\noindent which would correspond to a ``corrected'' form of equ.~(\ref{equ:no_wind}) for non-ideal plume sources. The variation of the interface height with respect to the source mass flux is plotted in Fig.~\ref{fig:xi_vs_qs_woods}.

\noindent The height of the interface is decreasing with the increasing mass flux due to the forcing of the heat mass flux. More air will be introduced inside the building and for a fixed effective area the volume taken by the buoyant layer will be larger. As a consequence the interface will tend to be pushed against the floor. It is interesting to see that for a certain threshold the interface height reaches the bottom of the building. This corresponds to the case where the heat source flux is very large (relative to the given effective area) and discharged plume fluid will completely fill the room. The other extreme case for when the interface height reaches the level of the ceiling corresponds to low heat source flux where the amount of air entering the building is too small to create a buoyant layer, which instead directly flows out through the (top) opening.
\newpage

\subsection{Full model with non ideal source} \label{sec:therory-full-non-ideal}
Finally, all the parameters ruling the flow are taken into account. In this scenario, the relation between the non-dimensional interface height, the pressure drop and the source mass flux can be obtained with a combination of equ.~(\ref{equ:wind}) and equ.~(\ref{equ:woods_modified}):

\begin{equation} \label{equ:full_equ}
\frac{A^*}{H^2} = \frac{ \lambda^{3/2} \left[ (\xi+\zeta_s)^{10/3} - \frac{A_T^2}{A_T^2 + A_B^2} \left( 2 \zeta_s^{5/3} (\xi+\zeta_s)^{5/3} -\zeta_s^{10/3} \right) \right]^{1/2} }{ \sqrt{ \frac{ 1-\xi }{ (\xi+\zeta_s)^{5/3} } - \lambda \frac{ \Delta p}{\rho_0} \left( \frac{H}{B_s} \right)^{2/3} } }
\end{equation}

\noindent This equation has not yet been validated by experimental data in the literature. By way of validation, the limiting cases for $\Delta p = 0$ (no wind) and $\zeta_s = 0$ (ideal source) have been solved analytically and equ.~(\ref{equ:full_equ}) is converging to equ.~(\ref{equ:no_wind}), respectively equ.~(\ref{equ:woods_modified}) which serves as a validation for those limiting cases. Even though the full validity has not been proven, since no other theoretical model has been found in the literature, this equation will be used for the comparison with the numerical simulations.

\begin{figure}[ht]
\centering
  \includegraphics[width=0.8\textwidth]{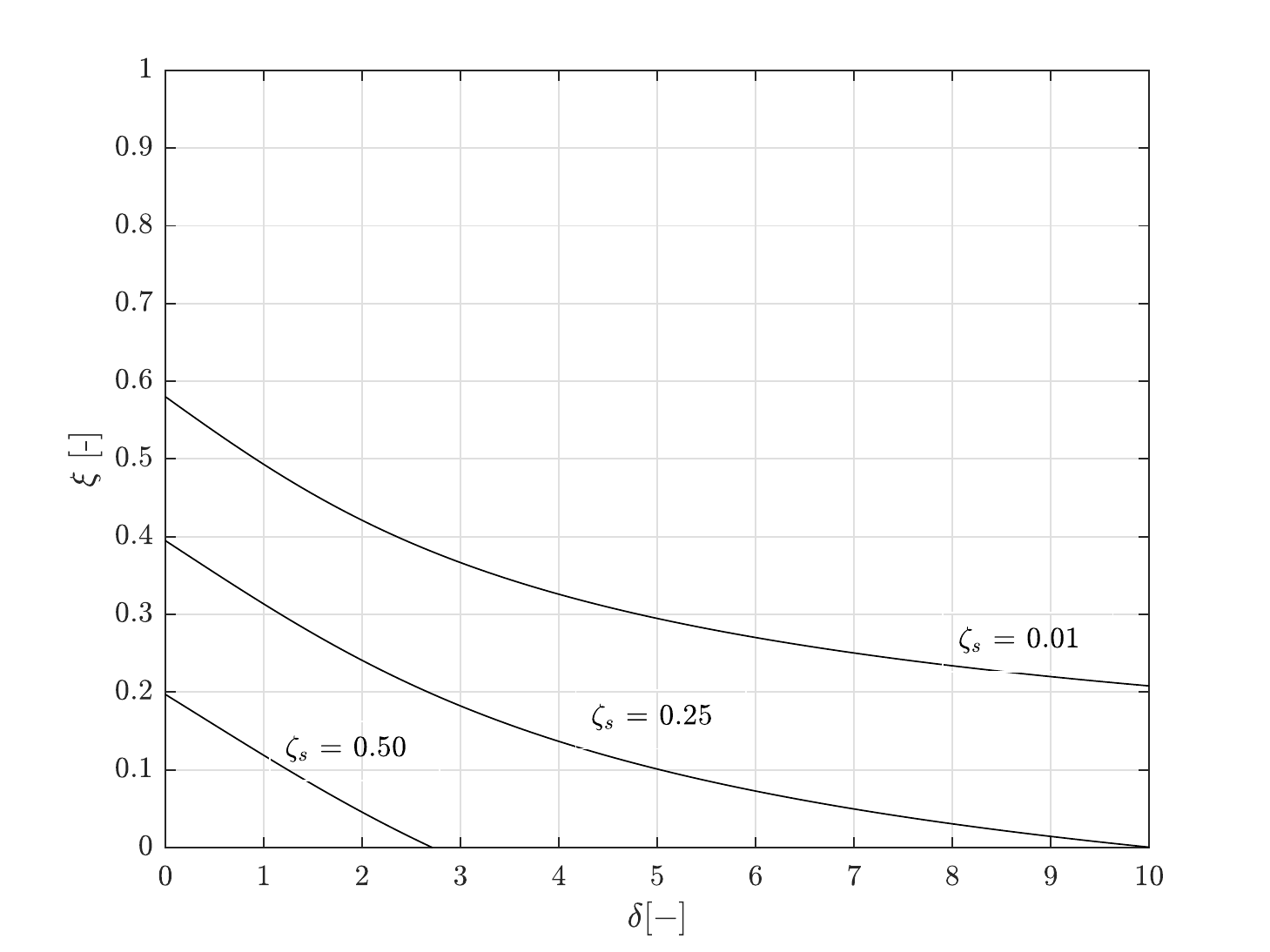}
  \caption{Variation of the non-dimensional interface height with respect to the non-dimensional parameter $\delta$ for different non ideal heat sources $\zeta_s$ and a normalized effective area of $A^*/H^2$ = 0.021.}
  \label{fig:xi_vs_delta_flynn}   
\end{figure}

\noindent Fig.~\ref{fig:xi_vs_delta_flynn} represents the variation of the non-dimensional interface height with regards to the non-dimensional parameter $\delta$ for different source mass fluxes. As expected the interface height is lower for higher pressure differences over the two sides of the building since the wind will ``push'' the interface against the bottom. On the other hand, as already seen in Fig.~\ref{fig:xi_vs_qs_woods} the lowering of the interface will be amplified by the strength of the source mass flux.\\

This time, the characteristic polynomials determining the relative importance of wind, the internal buoyancy respectively cannot be determined and expressed with one equation only, similarly to what was done for the case of an ideal source (equ.~\ref{equ:states1}). Three equations must be used, taking into account the source volume flux. To keep all the parameters non-dimensional the characteristic volume flux is chosen differently and instead of using $Q_w$ (as was done in the derivation of equ.(\ref{equ:states1}), but rather $Q_{B_{s}} = A^* B_s^{1/3} H^{1/3}$. Accordingly, the non-dimensional source volume flux and the non-dimensional flux through the top opening are defined as:

\begin{equation}
q_s = Q_s/Q_{B_{s}} \quad and \quad q_T = Q_T/Q_{B_{s}}
\end{equation}

\begin{figure}[ht]
\centering
  \includegraphics[width=\textwidth]{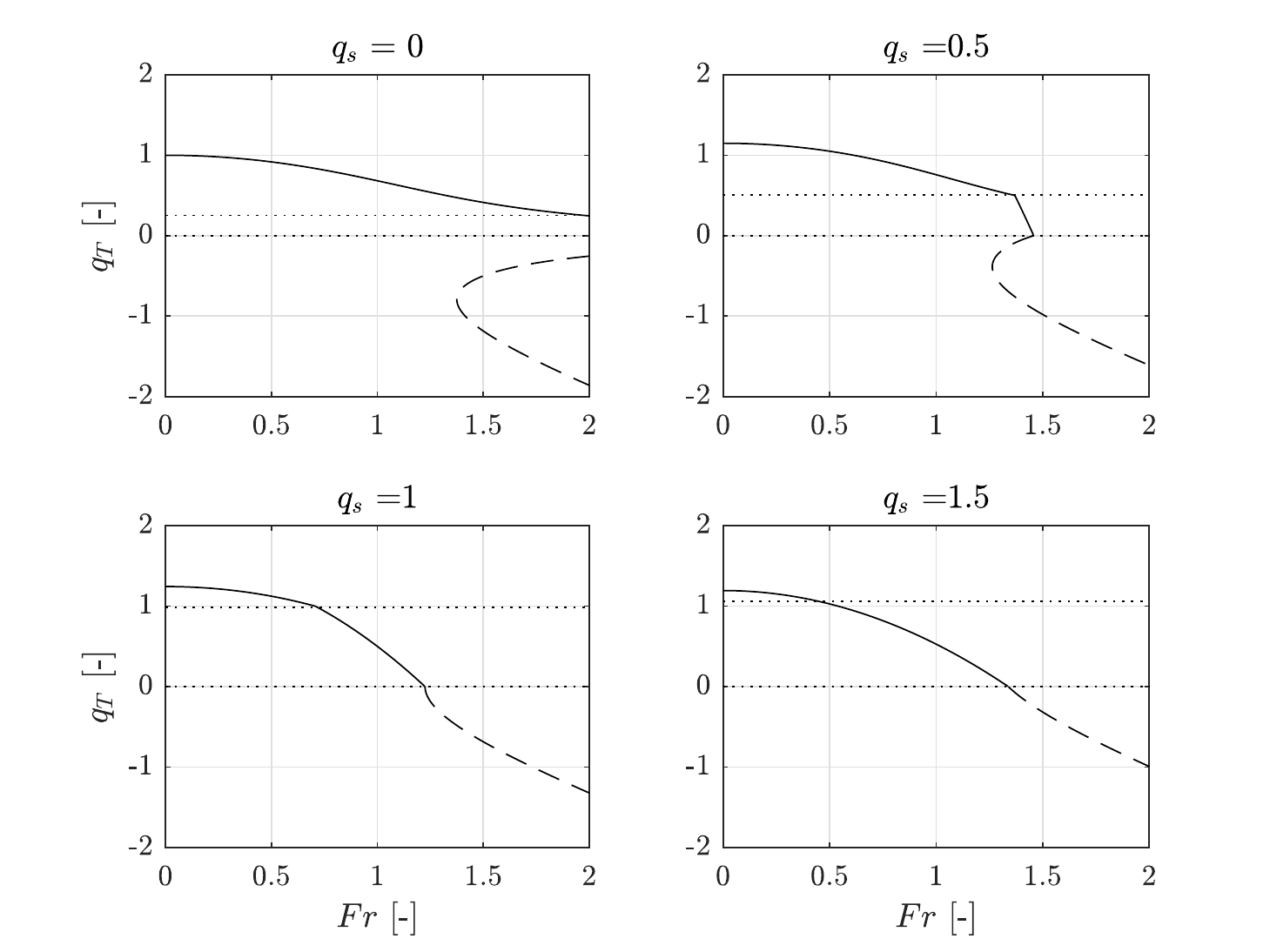}
  \caption{Non-dimensional flow rate through the top opening with respect to the Froude number for different source mass flux at a fixed effective area of $A^*/H = 0.021$.}
  \label{fig:qT_vs_Fr}   
\end{figure}

\noindent The three equations are corresponding to the three different regimes appearing in that kind of flow are:

\begin{equation} \label{equ:poly1}
q_T^{3} - 2(1-a^2)q_s q_T^2 + q_T \left[ (1-a^2) q_s^2 Fr^2 \right] - 1 +\xi = 0
\end{equation}
\begin{equation}\label{equ:poly2}
(1-2a^2)q_s q_T^2 - 2(1-a^2) q_s^2 q_T + (1-a^2)q_s^3 - q_s Fr^2 + 1 = 0
\end{equation}
\begin{equation}\label{equ:poly3}
q_T^3 - (3-2a^2) q_s q_T^2 + \left[ 3(1-a^2)q_s^2 -Fr^2 \right] q_T -(1-a^2)q_s^3 + q_s Fr^2 -1 = 0
\end{equation}

\noindent for a naturally-ventilated regime equ.~(\ref{equ:poly1}), blocked regime equ.~(\ref{equ:poly2}), and in the case of a well-mixed flow regime equ.~(\ref{equ:poly3}). The Fig.~\ref{fig:qT_vs_Fr} represents the resulting curves \cite{Flynn2009}, they are plotted for different non-dimensional source mass flux in function of the Froude number at a fixed effective area. The upper part of the solution (continuous line) corresponds to a stratified regime, the lower part (dashed line) to the mixed regime. The two dotted lines show the limits of the blocked regime (i.e. the solution in-between is blocked). The curves look different from the one presented in Fig.~\ref{fig:QQw_vs_Fr} because the normalization flux is different but the general idea is the same and they allow the prediction of the flow in a similar manner.

\subsection{Decoupled heat sources}
The heat source which was previously assumed to supply both heat and mass to the interior space is this time considered to supply heat and mass independently. This kind of model has so far never been studied analytically. The main objective of decoupling the heat source is to see if it involves any changes in the prediction of the interface height that might be useful to improve the existing models. Therefore, a numerical investigation only will be carried out to study any relevant influence of that modification on the final prediction.


\section{The numerical model}

\ifpdf
    \graphicspath{{Chapter3/Figs/Raster/}{Chapter3/Figs/PDF/}{Chapter3/Figs/}}
\else
    \graphicspath{{Chapter3/Figs/Vector/}{Chapter3/Figs/}}
\fi


\subsection{The finite volume method}

Within the framework of this project the commercial software ANSYS Fluent was used to solve the flows. This software is based on the finite volume method, which solves the partial differential equations (cf. equ.~\ref{equ:continuity}-3) describing the behavior of the flow iteratively, until a solution reasonably (and arbitrarily) close to reality is found. Theses equations are solved numerically on a mesh, constituted of small jointed volumes (or surfaces in 2D). 

\subsection{Governing equations}

In fluid mechanics, the Navier-Stokes equations are a set of non-linear partial differential equations describing the Newtonian flows behavior. The resolution of these equations by modeling the fluid as a continuum is extremely complex and analytical solutions only exist for a few simplified cases. Those equations are derived from the conservation laws applied to mass and momentum. In this case, the conservation of energy is added to the Navier-Stokes equations in order to model thermal effects \cite{Loomans1998}.\\

\textit{Conservation of mass (continuity equation):}
\begin{equation} \label{equ:continuity}
	\frac{\partial \rho }{\partial t } + \frac{\partial }{\partial x_i } (\rho u_i) =  0
\end{equation}

\textit{Conservation of momentum:}
\begin{equation} \label{equ:momentum}
    \frac{\partial}{\partial t }(\rho u_i) + \frac{\partial}{\partial x_j } ( \rho u_i u_j )  =  - \frac{\partial p}{\partial x_i } + \frac{\partial}{\partial x_j } \left[ \nu \left( \frac{\partial u_i }{\partial x_j } + \frac{\partial u_j}{\partial x_i } \right) \right] + \rho g_i 
\end{equation}

\textit{Conservation of energy:}
\begin{equation} \label{equ:energy}
\frac{\partial}{\partial t } (\rho H ) + \frac{\partial}{\partial x_i } (\rho u_i H ) = \frac{\partial}{\partial x_i } \left[ \frac{ K }{ c_p }  \frac{ \partial H }{ \partial x_i } \right] + S_H
\end{equation}

\noindent where $\rho$ is the density of the fluid, $u_i$ is the velocity component ($u,v,w$), $p$ is the pressure, $\mu$ indicates kinematic viscosity, $H$ the enthalpy and $S_H$ a source term. $K$ and $c_p$ represent the thermal conductivity, respectively the specific heat of the fluid. The time is indicated with $t$, $x_i$ is the coordinate axis (x,y,z) and $g_i$ is the gravitational acceleration.

\subsection{Turbulence modeling}

The Direct Numerical Simulation of these equations (DNS) requires a very fine mesh since the whole range of spatial and temporal scales of turbulence are resolved and the smallest eddies in the flow need to be captured. According to \citet{Nieuwstadt1990} the number of grid points required to describe turbulent motions with this method is at least $Re^{9/4}$ which would induce huge computational times (several years) for the case of naturally-ventilated buildings and are therefore not imaginable nowadays, even with the help of large computing clusters.\\

The second type of simulations, called Large Eddy Simulations (LES) is mostly based on the work of \citet{Smagorinsky1963} and \citet{Deardoff1970} proving the turbulent motion can be separated into large eddies and small eddies without having a significant impact on the evolution of the large eddies. Therefore, the large eddies are directly simulated whereas the small eddies are modeled with turbulent transport approximations. Thus, LES is clearly superior to turbulent transport closure wherein the transport terms are treated empirically. This techniques is becoming more and more relevant these days with the rapid increase in computational power on commercial machines.\\

Nonetheless, in the case of this study a Reynolds Averaged Navier-Stokes (RANS) model will be used both because of its computational efficiency and because of the broad application of RANS-based models in industry. To this end, the mean parameters are more useful than the instantaneous turbulent-flow parameters in the case of naturally ventilated buildings thus these models will give appreciable results \cite{Zhiqiang2007} that are of interest to architects, mechanical engineers and others interested in modeling transport processes in the built environment. These average-models for the turbulent flow have calculated the statistical characteristics of the turbulent motions by averaging the flow equations over time. The solution variables in the continuous Navier-Stokes equations are thereby decomposed into the mean and fluctuating components \cite{theory_guide}.

\begin{equation} \label{equ:vel_average}
	u = \overline{u}+u'
\end{equation}

\noindent with $\overline{u}$ and $u'$ being the mean and fluctuating components of the velocity field. The other variables (such as pressure and scalar quantities) are also decomposed same way.

\begin{equation} \label{equ:scal_average}
	\phi = \overline{\phi}+\phi'
\end{equation}

\noindent By introducing equ.~(\ref{equ:vel_average}) and equ.~(\ref{equ:scal_average})  into the continuous equ.~(\ref{equ:continuity} to 3.3) and taking a time average (ensemble average), new non-linear terms appear in the set of equations:

\begin{equation} 
	- \rho \overline{u_i' u_j'} \quad and \quad - \rho \overline{u_i' H'} 
\end{equation}

\noindent With the apparition of these new terms (Reynolds terms), the set of equations introduces a closure problem and more equations are necessary to solve it. Thus, the effect of turbulence is represented as an increased viscosity/diffusivity by its replacement with linear terms and scalar coefficients based on the  hypothesis (\citet{Boussinesq1887}) that the turbulent stresses are proportional to the mean velocity gradients.


\noindent These last non-linear fluctuating terms need to be modeled in order to obtain computationally solvable forms of the RANS equations. Numerous models have been developed over the past years and even though they will all - in general - give a solution, their ability to correctly predict the behavior of a flow may vary significantly. Therefore, choosing the right model for turbulence is a key element to the success of numerical flow simulations.\\

Many researchers have been trying to assess different turbulent models for naturally ventilated buildings and the two most popular RANS models that have been  successfully used for indoor ventilation flows are SST $k- \omega$ and RNG $k-\epsilon$ \cite{Zhiqiang2007}. They showed the best overall performance compared to the other models in terms of accuracy, computational efficiency and robustness. Moreover, the RNG $k-\epsilon$ has proven to be superior to the SST $k-\omega$ for low-turbulence flow \cite{ Craven2006, Zhang2005} such as the ones studied in this thesis. For these reasons, the theoretical development of the RNG $k-\epsilon$ model only will be made in details.\\

The standard $k-\epsilon$ model is based on model transport equations for the turbulent kinetic energy $k$ and its dissipation rate $\epsilon$. The derivation of the model is based on the assumption that the flow is fully turbulent and that the effects of molecular viscosity are negligible \cite{theory_guide}. It is therefore only valid for fully turbulent flows and would fail to predict the majority of low-turbulence flow appearing in natural ventilation. For these reasons, the renormalization group theory (RNG) is used. It has additional terms in the $\epsilon$ equation which provide analytically derived differential formula for the effective viscosity that accounts low-Reynolds number effects. The transport equations (\ref{equ:continuity}-3.3) are rewritten in function of the turbulent kinetic energy $k$ and the dissipation rate $\epsilon$:
\begin{equation}
\frac{\partial (\rho k)}{\partial t } + \frac{\partial }{\partial x_i } (\rho k u_i) =  \frac{\partial}{\partial x_j} \left[ \left( \nu + \frac{\nu_t}{\sigma_k} \right) \frac{\partial k}{\partial x_j} \right] + G_k + G_b - \rho \epsilon - Y_M + S_k
\end{equation}
and
\begin{equation} \label{equ:RNG_ke_2}
\frac{\partial (\rho \epsilon)}{\partial t } + \frac{\partial }{\partial x_i } (\rho \epsilon u_i) =  \frac{\partial}{\partial x_j} \left[ \left( \nu + \frac{\nu_t}{\sigma_{\epsilon}} \right) \frac{\partial \epsilon}{\partial x_j} \right] + C_{1 \epsilon} \frac{\epsilon}{k} (G_k + C_{3 \epsilon}G_b) - C_{2 \epsilon} \rho \frac{\epsilon^2}{k} + S_{\epsilon}
\end{equation}

\noindent $G_k$ represents the generation of turbulent kinetic energy due to the mean velocity gradients, $G_b$ is the generation of turbulent kinetic energy due to buoyancy. $Y_M$ represents the contribution of the fluctuating dilatation in compressible turbulence to the overall dissipation rate. $C_{1 \epsilon}$, $C_{2 \epsilon}$ are constant values determined analytically by the RNG theory, $C_{3 \epsilon}$ is also a constant whose determination will be explained below, $\sigma_k$ and $\sigma_{\epsilon}$ are the turbulent Prandtl numbers for $k$ and $\epsilon$. $S_{k}$ and $S_{\epsilon}$ are the source terms that account for the production of kinetic energy, respectively dissipation rate at the simulated sources.\\

As it was already mentioned earlier, the effect of buoyancy is one of the major flow-driving aspects of naturally ventilated buildings and it is therefore important to understand the way its modeled. For $k-\epsilon$ the generation of turbulence models is given by
\begin{equation}
G_b = \beta g_i \frac{\mu_t}{Pr_t} \frac{\partial T}{\partial x_i}
\end{equation}
where $Pr_t$ is the turbulent Prandtl number for energy and $g_i$ the component of the gravity in the associated direction. For the standard $k-\epsilon$ model the Prandtl number is considered constant but in the case of the RNG model it is computed using a formula analytically derived by the renormalization group theory to give it more accuracy.\\

In ANSYS Fluent the buoyancy effects on the generation of $k$ are relatively well understood \cite{user_guide} but its effects on $\epsilon$ are less clear. By default, they are neglected by keeping the term $G_b$ in the transport equation (\ref{equ:RNG_ke_2}) to zero. However, recent versions of ANSYS offer the possibility to include these effects in the solver's viscous model box. The degree of influence is weighted by the constant $C_{3 \epsilon}$, computed accordingly to $C_{3 \epsilon} = tanh|v/u|$ where $v$ is the component of the flow velocity  parallel to gravity and $u$ the component of the flow velocity perpendicular to $v$. In this way, the constant $C_{3 \epsilon}$ will be maximized (equal to 1) if the shear layer of the buoyant flow is aligned with gravity and zero if the buoyant shear layer is perpendicular, which will annihilate the influence of the buoyancy on the dissipation rate. Since the addition of this last parameter is fairly recent to Fluent (version 17.2) no proof of its efficiency has been found in the literature but it will be used accordingly to the recommendations of the user's guide \cite{user_guide}.

\subsubsection{Hydrostatic head and Boussinesq approximation}
Since gravitational acceleration is activated in the solver in order to model the buoyant effects, the pressure field will include the hydrostatic head. Mathematically, this is accomplished by redefining the pressure in terms of a modified pressure that includes the hydrostatic head (denoted p') as follow :

\begin{equation}
p' = p- \rho_0 g_i \cdot r_i
\end{equation}

\noindent where $r_i$ is the position vector component. Noting that 

\begin{equation}
\frac{\partial}{\partial x_i} (\rho_0 g_i r_i) = \rho_0 g_i
\end{equation}

\noindent it follows that

\begin{equation} \label{equ:nabla_p_prime}
\frac{\partial p'}{\partial x_i} = \frac{\partial}{\partial x_i} (p - \rho_0 g_i \cdot r_i) = \frac{\partial p}{\partial x_i} - \rho_0 g_i
\end{equation}

\noindent The substitution of equ.~(\ref{equ:nabla_p_prime}) in the momentum equation (equ.~\ref{equ:momentum}) gives a pressure gradient and gravitational body force terms of the form

\begin{equation}
-\frac{\partial p'}{\partial x_i} + (\rho - \rho_0) g_i
\end{equation}

\noindent where $\rho$ is the actual fluid density. Since the Boussinesq hypothesis is made, the body force term becomes

\begin{equation}
(\rho - \rho_0) g_i \approx \rho_0 \beta (T-T_0) g_i
\end{equation}

\noindent The consequence of this treatment of the gravitational body force is that the report of static and total pressures will not show any influence of the hydrostatic pressure. It is important to note this is consistent with equ.~(\ref{equ:wind}) proposed by \citet{Linden1999}, \citet{Flynn2009}. Their models do not take into account the hydrostatic head over the building so that the position of the openings on the sides of the building should not influence the prediction of the interface height. The advantage of using the Boussinesq hypothesis over a compressible gas relation is that it simplifies considerably the equations solved since the density is considered constant for all terms except when it is multiplied by gravity. This allows faster convergence thus shorter simulation times. The hypothesis is accurate as long as changes in actual density are small, specifically, it is valid when $\beta(T-T_0) \ll 1$ which is the case in the current setup (thermal expansion coefficient of air: $\beta$ = 0.0034 K$^{-1}$, maximum temperature difference: $\Delta T$ = 10 K).


\section{Numerical simulation of the naturally ventilated flow}

\subsection{Simulation of the emptying-filling box}
\subsubsection{Description of the problem}
\begin{figure}[ht]
\centering
  \includegraphics[width=0.5\textwidth]{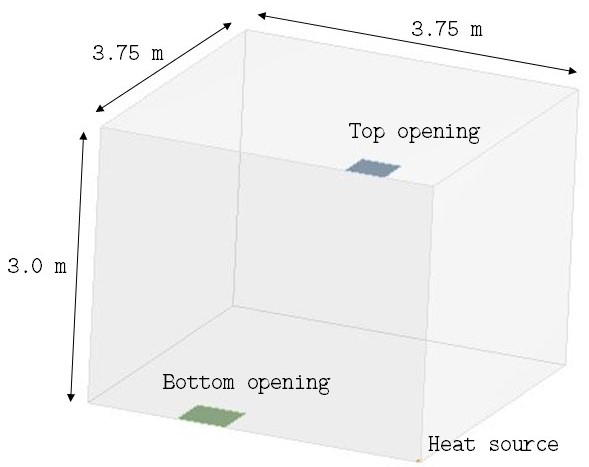}
  \caption{Representation of the numerical domain used for the simulations of the building with an ideal heat source only.}
  \label{fig:slide6}   
\end{figure}

The first set of numerical simulations that were carried out aimed at simulating the effect of a buoyancy driven flow inside a single-zone building. To validate the developed model, the numerical simulations performed by \citet{Kaye2009} were reproduced. The domain considered is a 7.5m $\times$ 7.5m  $\times$ 3m building as it can be seen in Fig.~\ref{fig:slide6}.

\noindent Only one-quarter of the domain was simulated due to the symmetry of the situation, allowing to reduce the computational times necessary for the simulations. Different effective areas for openings located on the ceiling and the floor were studied as well as different source powers. These different parameters are presented in the Table \ref{tab:Kaye2009}.

\begin{table}[ht]
\centering
\caption{Different parameters used for the simulations with openings on the ceiling/floor and no wind}
\label{tab:Kaye2009}
\begin{tabular}{c c r r r r }
	\multicolumn{2}{l}{Case} & $A^*/H$ & P [W]& $\mu$ & $\sqrt{t_d / t_f}$\\
	1 & $\quad$ & 0.0013 & 3.20 & 38.2 & 27'463   \\   
    2 &			& 0.0013 & 320.00 & 38.2 & 5'917 \\     
    3 &			& 0.0053 & 0.08 & 9.5 & 46'962  \\    
	4 & 		& 0.0104 & 3.20 & 4.9 & 9'825   \\   
	5 & 		& 0.0319 & 3.20 & 1.6 & 5'612  \\    
	6 & 		& 0.0628 & 3.20 & 0.8 & 4'003 \\   
	7 & 		& 0.0840 & 80 & 0.6 & 1'834  \\ 
    
\end{tabular}
 \end{table}

\subsubsection{Mesh topology} \label{sec:ideal-no-wind-mesh}
\begin{figure}[ht]
\centering
  \includegraphics[width=0.8\textwidth]{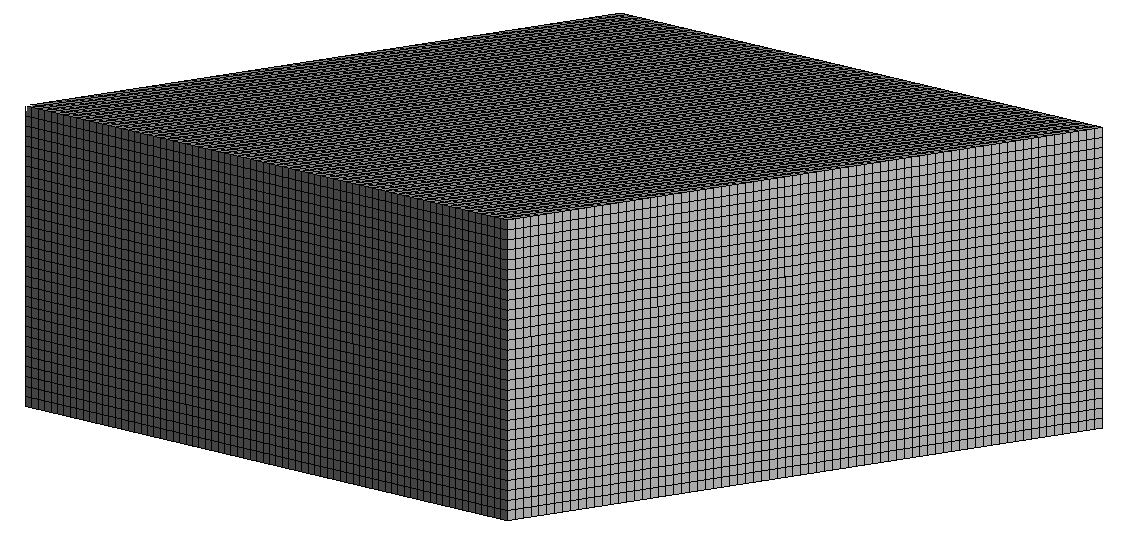}
  \caption{Illustration of the mesh used for the simulation of the building with an ideal heat source only.}
  \label{fig:building_mesh}   
\end{figure}

Regarding the squareness of the domain, a mesh constituted of hexahedron was generated. It is the simplest and fastest way to mesh such a domain. For information purposes, the mesh used is pictured in Fig.~\ref{fig:building_mesh} below even though it is very straightforward to imagine what such mesh would look like (with symmetry).

\noindent In order to optimize the number of cells used for the simulations by minimizing the computational times without loosing in accuracy and robustness, a convergence study was carried out. It was realized by comparing the interface height for different meshes for a normalized effective area of $A^*/H^2$ = 0.084 and a heat source of $P$ = 80 W. The mesh was considered robust when the interface height did not change, regardless of the reduced number of cells. A mesh with $\sim$ 90'000 cells proved to give a robust and accurate solution and was therefore kept for the whole set of simulations.

\subsubsection{Operating conditions}
The operating pressure is set to $P_0$ = 101325 Pa which is the default pressure in Fluent. Gravity is activated and the reference density and temperatures are set to $\rho_0$ = 1.225 kg/m$^3$, respectively $T_0$ = 293 K.

\subsubsection{Boundary conditions} \label{sec:bc-ideal-no-wind}
The boundary conditions were defined as follow:
\begin{itemize}
	\item \textbf{\textit{bottom opening:}} pressure inlet with 0 Pa gauge pressure, backflow temperature of $T_0$ and with very low level of turbulence ($k$ = 10$^{-5}$ and $\epsilon$ = 10$^{-6}$)
    \item \textbf{\textit{top opening:}} pressure outlet with 0 Pa gauge pressure, backflow temperature $T_0$ and very low level of turbulence ($k$ = 10$^{-5}$ and $\epsilon$ = 10$^{-6}$) 
    \item \textbf{\textit{heat source:}} the heat source is modeled as a heating plate producing a heat flux corresponding to the heat outputs $P$ shown in Table \ref{tab:Kaye2009}.
    \item \textbf{\textit{building walls, ceiling and floor:}} the walls and the floor are considered adiabatic with a standard no slip condition.
\end{itemize}

\subsubsection{Discretization scheme}
All the discretization schemes (Pressure, momentum, turbulent kinetic energy, turbulent dissipation rate and energy) were chosen of the second order. A SIMPLE scheme was used for the pressure-velocity coupling.
 
\subsubsection{Wall treatment}
\begin{figure}[ht]
\centering
  \includegraphics[width=\textwidth]{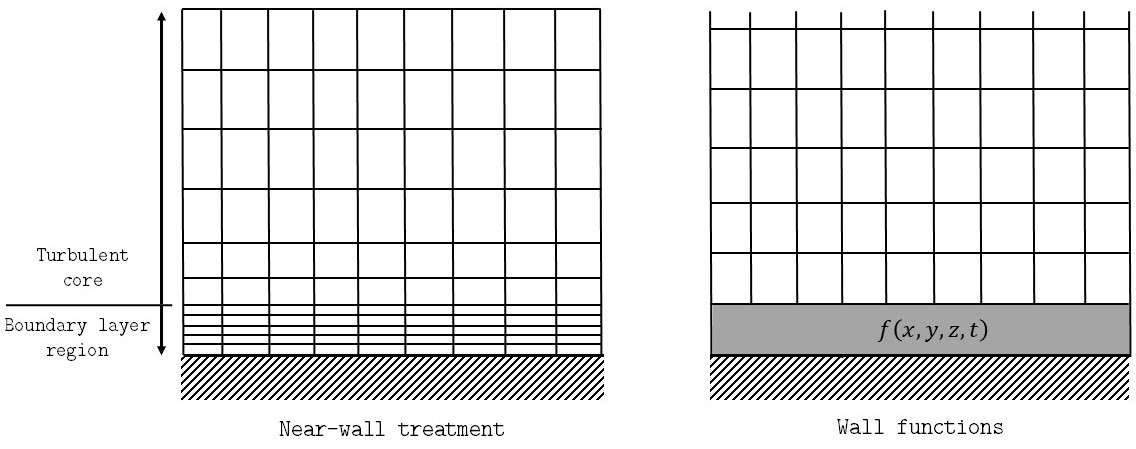}
  \caption{Representation of the two main techniques to model boundary layers regions: near-wall treatment (left) and wall functions (right).}
  \label{fig:slide7}   
\end{figure}

Two approaches are used for the modeling of boundary layers: near-wall treatment and wall functions. By nature, these two approaches require different types and sizes of cells. The first one consists of having a mesh that is fine enough in the boundary layer region to accurately model the physics happening there with the chosen turbulence model, as it is shown on the left of Fig.~\ref{fig:slide7}. On the other hand, the  wall functions approach ``bridges'' the near wall region with experimentally validated function that reproduce the boundary layer and the associated logarithmic profile and heat transfer phenomena (right of Fig.~\ref{fig:slide7}). The advantage of the second method is that it allows to obtain robust and accurate results without requiring a very fine mesh and was therefore chosen in this case.

\subsubsection{Time-stepping}
The management of the time-steps was inspired by the simulations carried out by \citet{Kaye2009}. Smaller time-steps with a large number of iterations were used during the early stages of the flow when the flow field is establishing and highly variable gradients exist. Moreover once the flow field is established, larger time-steps were used with larger number of iterations, accordingly to the procedure presented by \citet{Kaye2009}: 1 second for the first 10 time steps, 5 seconds for the time steps 11-20, 10 seconds for time steps 21-30, followed by 42 time steps at 20 seconds each. After those initial 1000s fixed time steps of 50 seconds with 50 maximum iterations were used to achieve the final steady state.

\subsubsection{Results}
\begin{figure}[ht]
\centering
  \includegraphics[width=\textwidth]{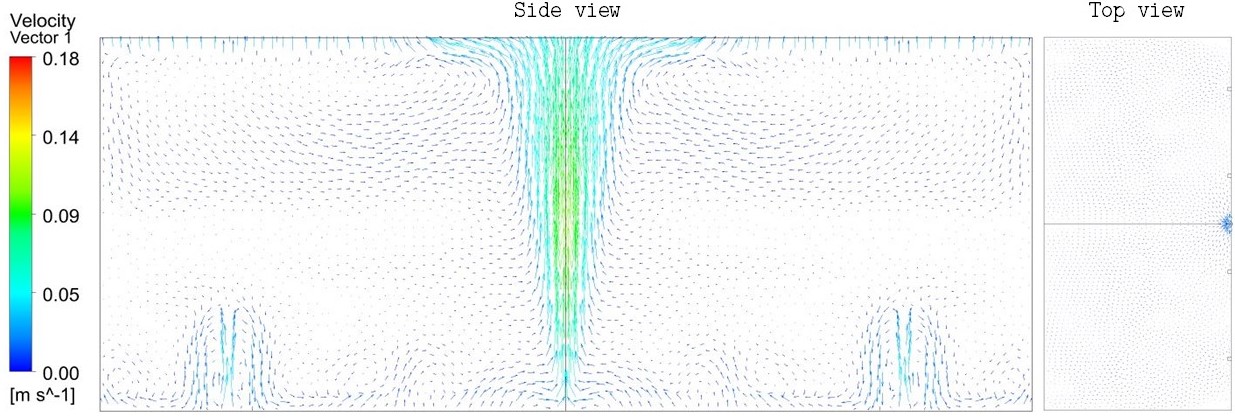}
  \caption{Velocity field inside the simulated building for a fixed effective area of $A^*/H^2$ = 0.003 and a heat source power of 80W. The side view (left) is plotted on a plane parallel to the wall at the center. The top view (right) is plotted at the height of the density interface. Symmetry is applied for both images.}
  \label{fig:standard-openings-velocity-field}   
\end{figure}

\textbf{Velocity Field}\\
The velocity field inside the building is shown in Fig.~\ref{fig:standard-openings-velocity-field} on a plane parallel to the wall at the center of the plume. A typical plume profile can be observed with high vertical speeds in the center and low speed regions with recirculation on the edges. The radius of the plume is increased with the height due the entrainment of more air with buoyancy forces. The bottom openings are providing air to the system and it can clearly be seen in the velocity field by the two ``bumps'' on the sides. This field is similar to what was obtained (and validated) by \citet{Kaye2009} and is therefore assumed correct.

\noindent Looking at the right part of Fig.~\ref{fig:standard-openings-velocity-field} it is interesting to note that the flow is almost at rest in the part of the building away from the center of the plume. The turbulent flow is confined in the plume region. The main reason for this is that the top openings of the building are located nearby the center of the room and the turbulent flow created by the plume can quickly escape the building without disturbing the rest of the flow.\\ 

\begin{figure}[ht]
\centering
  \includegraphics[width=\textwidth]{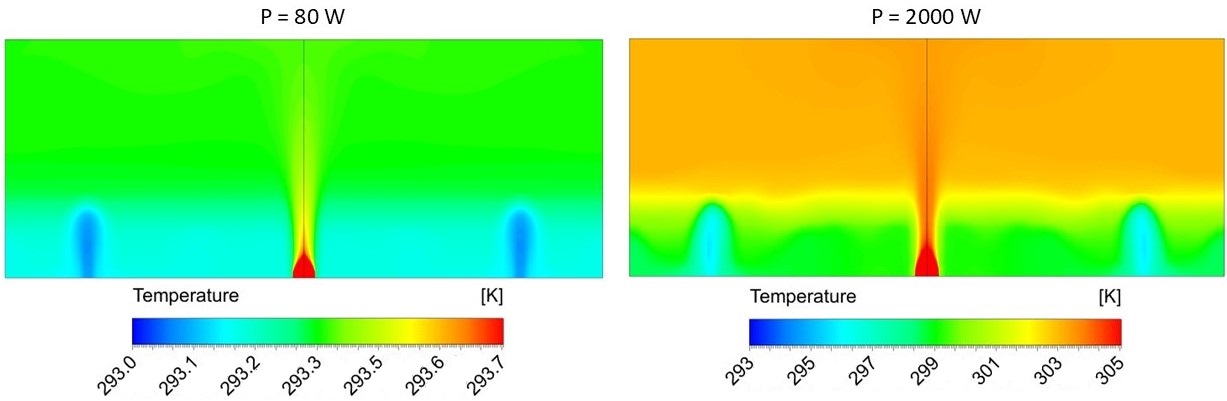}
  \caption{Temperature field inside the simulated building for a fixed effective area of $A^*/H^2$ = 0.003 plotted on a plane parallel to the wall at 0.\,m off the center. The left part corresponds to a heat source power of 80\,W whereas the right one corresponds to the temperature field for a heat source of 2000\,W.}
\label{fig:standard-openings-temp-field}
\end{figure}

\textbf{Temperature field}\\
The temperature field inside the domain is plotted in Fig.~\ref{fig:standard-openings-temp-field} on a plane parallel to the wall at the center of the room for two different heat source powers. The plume is once again appearing clearly in the center of the room, with its radius increasing with height. It can be seen that the bottom openings are providing cold air to the interior space. The density interface is observed in the middle of the domain.

\noindent It is interesting to note that the density interface height is the same, for both cases the height of the interface is only depending on the effective area $A^*$. Those results agree with the model presented in theory and equ.~(\ref{equ:no_wind}). The main difference is seen in the transient behavior of the flow where stronger heat sources converge more quickly to a warmer steady state. The second difference is that the interface is more diffuse in the case of strong heat sources because of the stronger temperature gradients in the room and the overall higher degree of turbulence. The detailed impact of different heat source power was not studied in detail due the limited allocated time, but the results presented above show that the thermal effects of the flow should also be modeled correctly.\\

\begin{figure}[ht]
\centering
  \includegraphics[width=0.8\textwidth]{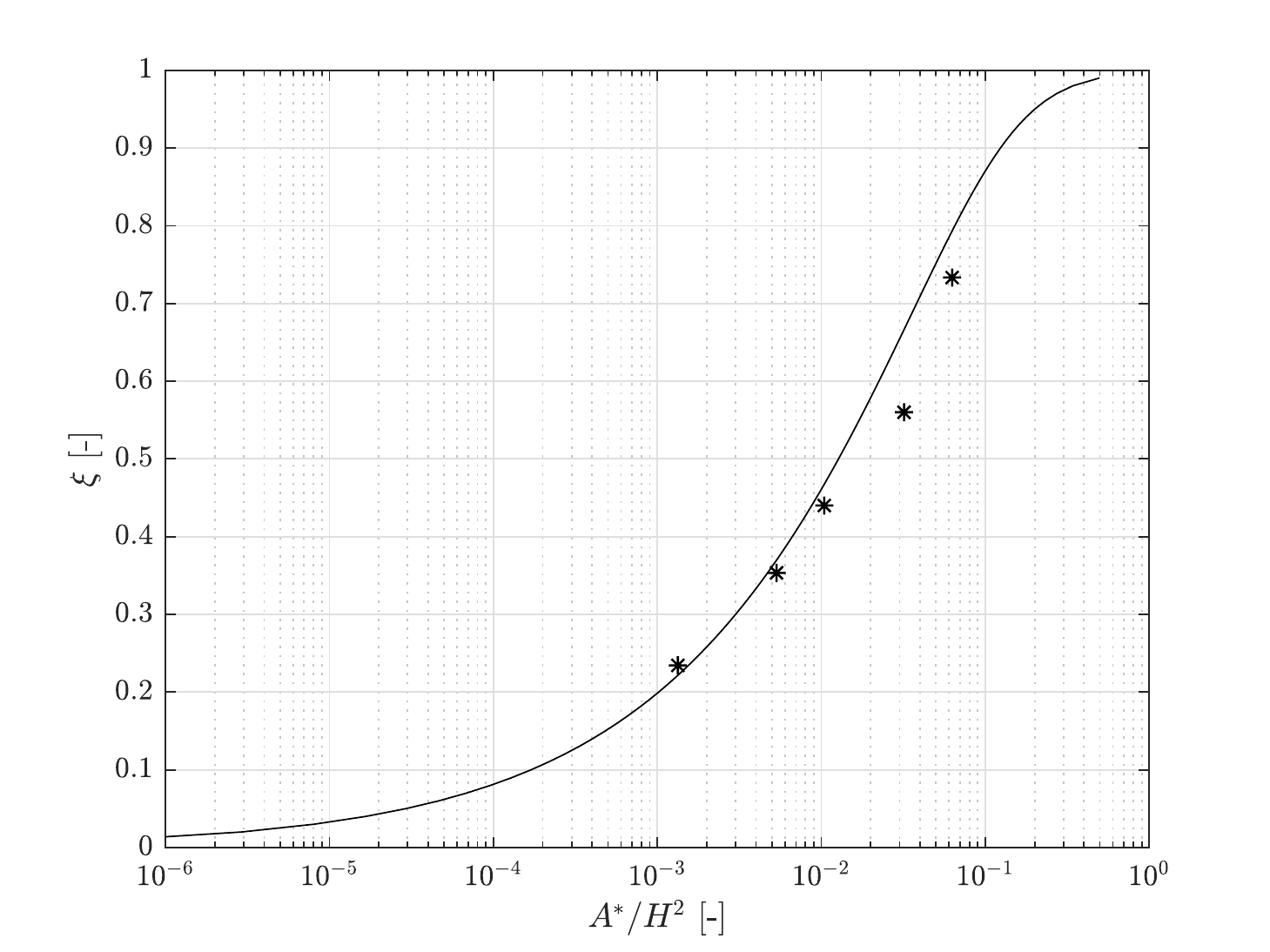}
  \caption{Variation of the non-dimensional interface height with respect the normalized effective area for an ideal heat source.}
  \label{fig:xi_vs_AH2_cfd}   
\end{figure}

\textbf{Interface height}\\
The interface height simulated with Fluent was compared to the theoretical (validated) value presented in section \ref{sec:ideal-no-wind-theo}. The evaluation of the numerical interface height is explained in \ref{app:measurements}. The results are displayed in Fig.~\ref{fig:xi_vs_AH2_cfd}. It shows a good agreement with the theoretical model over the whole range of simulated values. Different heat source power levels were tested but they did not change the final steady state value, only the transient behavior, i.e. higher heat source powers only gave faster convergence to the final steady state. These results agree with the numerical simulations produced by \citet{Kaye2009} and show that the mesh, turbulence model, discretization schemes and different parameters used for the simulations are able to predict the behavior of natural ventilation flows.\\

\begin{figure}[ht]
\centering
  \includegraphics[width=0.8\textwidth]{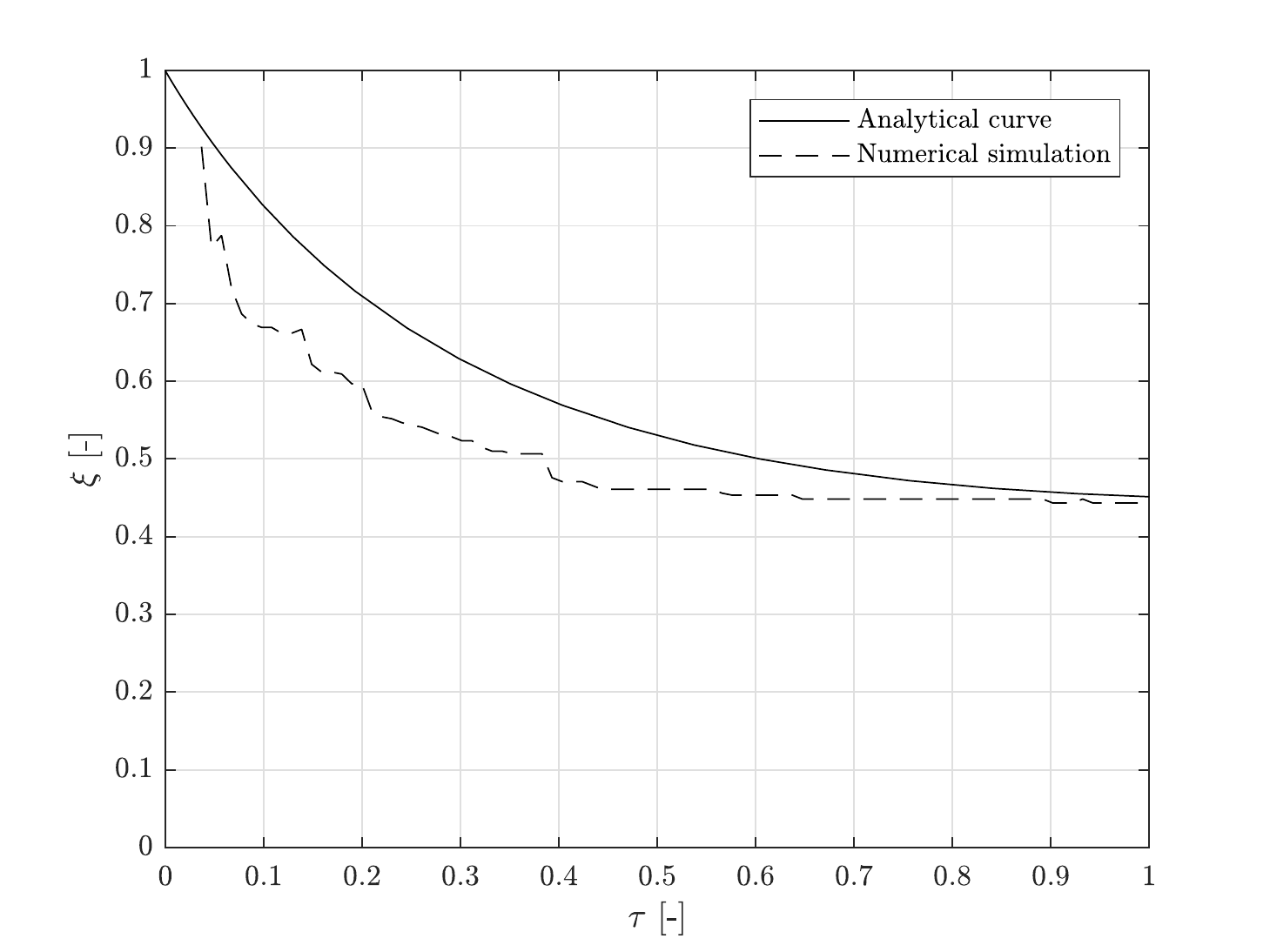}
  \caption{Evolution of the non-dimensional interface height with respect to the non-dimensional time $\tau$ for a normalized effective area of $A^*/H^2$ = 0.005 and a heat source power of 80\,mW.}
  \label{fig:xi_vs_tau_cfd}   
\end{figure}

\textbf{Transient behavior}\\
The transient behavior of the numerical flow was compared with the theoretical model presented in section \ref{sec:ideal-no-wind-theo} since these results had also a validation. This was done to verify that the numerical parameters used were performing well enough in reproducing the development of the flow and not only its final state. The transient behavior of the interface height was plotted for case 3 (see Table \ref{tab:Kaye2009}) and is represented in Fig.~\ref{fig:xi_vs_tau_cfd}. The behavior seems to be correctly modeled even though the value is slightly under-predicted before the steady state is reached.

\subsection{Simulation of the emptying-filling box with lateral openings}
\subsubsection{Description of the problem}
\begin{figure}[ht]
\centering
  \includegraphics[width=0.65\textwidth]{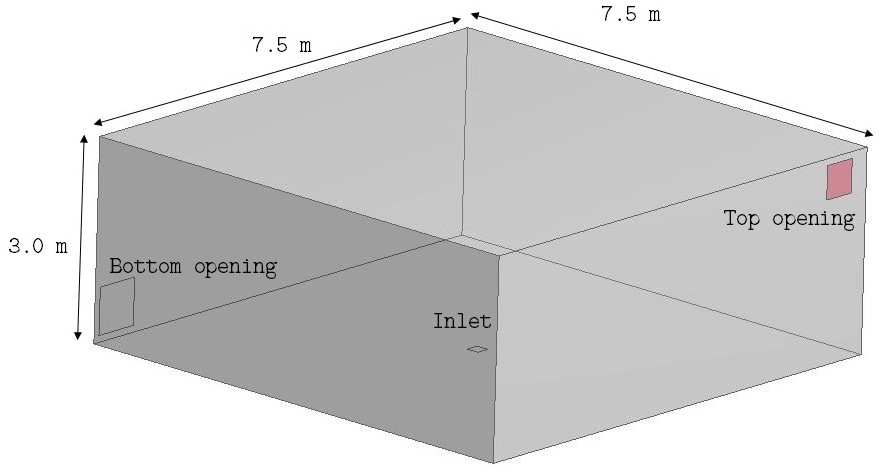}
  \caption{Representation of the numerical domain used for the simulation taking in account the influence of the external atmosphere.}
  \label{fig:building_domain}   
\end{figure}

This first set of simulations was carried out with the setup mentioned above since it was the closest (experimentally validated) study to the final objective of this project found in the literature. Since the further simulations where studying the influence of the external atmosphere and to account for the effect of the wind, it was not possible to put the openings of the building on the ceiling, respectively on the floor, but rather on the sides of the building (see Fig.~\ref{fig:building_domain}). Therefore, the simulations where re-run with the same mesh, turbulence model and parameters inputs, but solely changing the locations of the openings to the sides of the building. To asses this effect six cases were studied. The parameters for the different cases are presented in Table \ref{tab:lat-openings}. 

\begin{table}[ht]
\centering
\caption{Different parameters used for the simulations with lateral openings and no wind}
\label{tab:lat-openings}
\begin{tabular}{l c c c  }
	Case & $A_b$ [m$^2$] & $A_t$ [m$^2$] & $A^*/H^2$  \\
    \hline
	1 & 0.04 & 0.04 & 0.003  \\   
    2 & 0.09 & 0.09 & 0.006  \\     
    3 & 0.16 & 0.16 & 0.011  \\    
	4 & 0.25 & 0.49 & 0.021  \\   
	5 & 0.49 & 0.49 & 0.033  \\      
    \hline
\end{tabular}
\end{table}

\subsubsection{Results}
\begin{figure}[ht]
\centering
  \includegraphics[width=\textwidth]{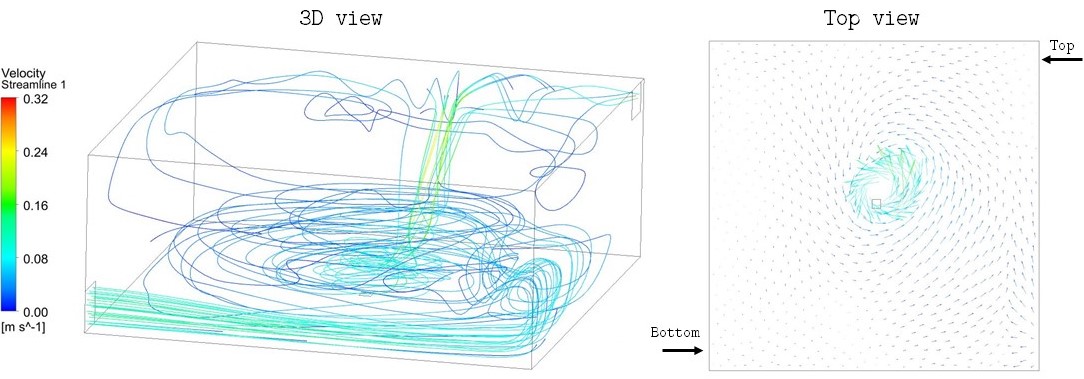}
  \caption{Streamlines starting from the bottom opening of the building (left) and the velocity field on the density interface (right) for an effective area of $A^*/H^2$ = 0.021 and an ideal heat source of 200\,W.}
  \label{fig:lat-openings-velocity-field}   
\end{figure}

\textbf{Velocity Field}\\
The streamlines starting from the bottom opening of the building as well as the velocity field at the interface height are plotted in Fig.~\ref{fig:lat-openings-velocity-field}. The flow exhibits much more turbulence than the one presented in Fig.~\ref{fig:standard-openings-velocity-field}. It has gained one more degree of freedom in the sense that when the openings were located on the floor/ceiling the air exchange with the external atmosphere was in the same direction with the plume whereas now air comes laterally. This induces a high vorticity in the flow that can be observed on the right part of Fig.~\ref{fig:lat-openings-velocity-field}.

\noindent Despite these phenomena, stratification still takes place and a buoyant layer is created. The interface is however less well-defined than for openings situated on the floor/ceiling, due to the higher levels of turbulence. Furthermore, the vortex is not steady and even though the plume keeps its conic shape it precesses around the center of the heat source. All these reasons explain why the interface height simulated is under-predicted compared to the theoretical model. A way of correcting this will be discussed in the following paragraphs.

\begin{figure}[ht]
\centering
  \includegraphics[width=\textwidth]{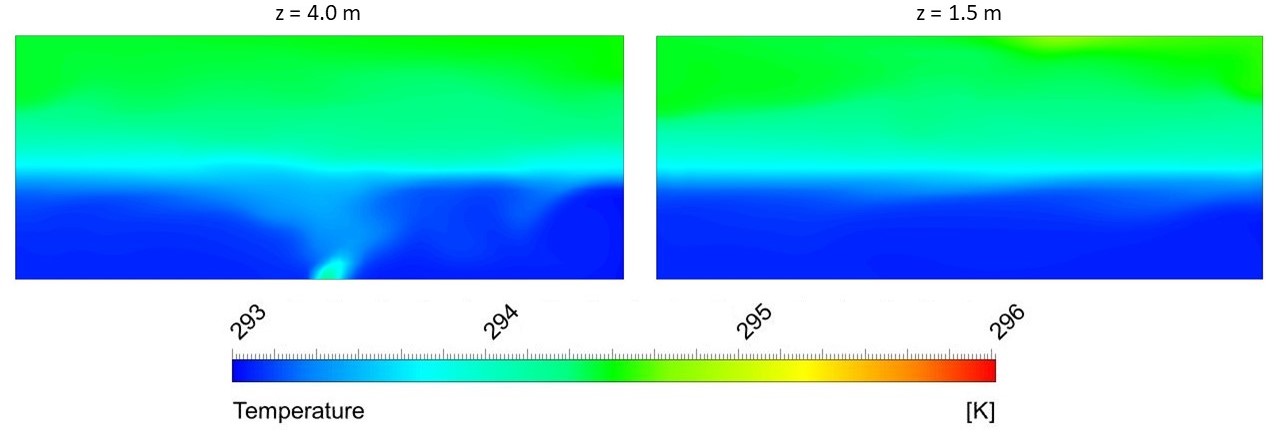}
  \caption{Temperature field inside the simulated building for a fixed effective area of $A^*/H^2$ = 0.021 plotted on a plane parallel to the wall at 0.3\,m off the center (left), respectively 1.5\,m (right).}
  \label{fig:lat-openings-temp-field}   
\end{figure}

\textbf{Temperature Field}\\
The temperature fields inside the domain are plotted in Fig.~\ref{fig:lat-openings-temp-field} on a plane parallel to the wall at 0.3 m and 1.5 m off the center of the room for two different heat source powers. Once again the density layer is clearly visible.

\noindent Similarly to what was said for the velocity field in the previous chapter, the interface is not as defined as for openings situated on the floor/ceiling of the building. Moreover, the more turbulent flow induces a mechanical diffusion of the heat through the interface. The precession of the plume can be seen on the left the part of Fig.~\ref{fig:lat-openings-temp-field} as the plume is not aligned with the vertical direction but slightly leans to the right.\\

\begin{figure}[ht]
\centering
  \includegraphics[width=0.8\textwidth]{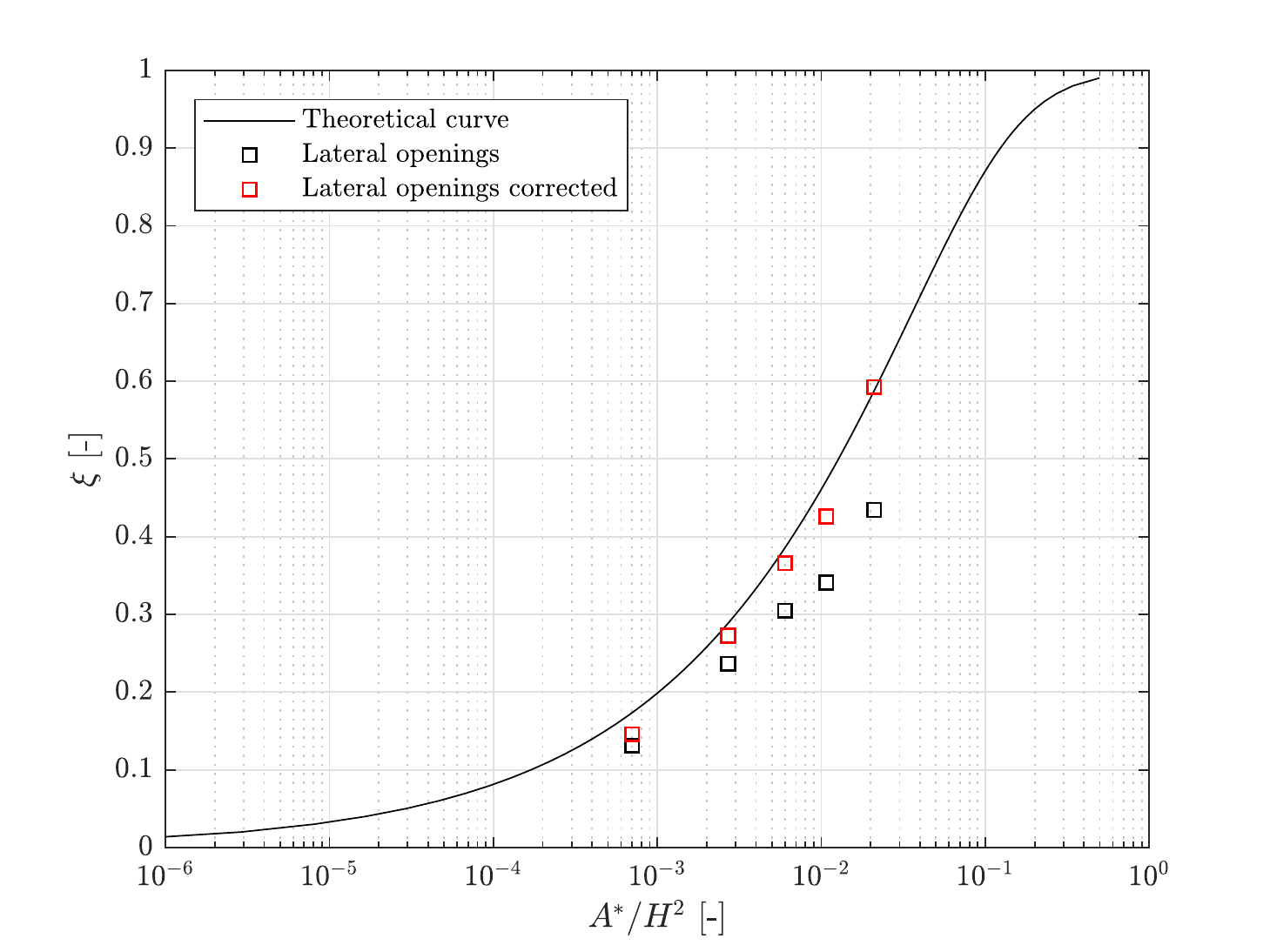}
  \caption{Variation of the non-dimensional interface height with respect the normalized effective area for an ideal heat source and lateral openings.}
  \label{fig:xi_vs_AH2_cfd_lat}   
\end{figure}

\textbf{Interface height}\\
\noindent Fig.~\ref{fig:xi_vs_AH2_cfd_lat} shows the results for this case, compared to the ones already presented before. It appears that the interface height is lowered for openings located on the sides rather than on the ceiling and bottom of the building respectively. This phenomenon can be explained by the influence of the openings on the total area of the walls that will represent a non-negligible percentage of the total height of the building.\\

\begin{figure}[ht]
\centering
  \includegraphics[width=0.5\textwidth]{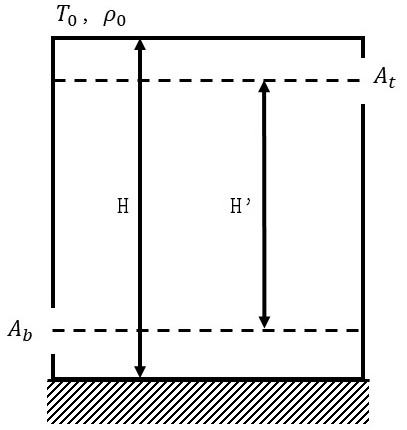}
  \caption{Illustration of the lateral view of the building. $H'$ is the corrected building height, taken from the middle of the openings to take into account their effect on the flow.}
  \label{fig:h_hprime}   
\end{figure}

\noindent The numerical model can be corrected by normalizing the interface height by the corrected height of the building $H'$. The corrected height is taken as the height between the middle of the openings as shown in Fig.~\ref{fig:h_hprime}. The corrected results are represented in red in Fig.~\ref{fig:xi_vs_AH2_cfd_lat} and seem to work fairly well for the whole range of simulations.\\

Regarding the analysis presented above, it would have been better to produce a correction model taking into account the vorticity of the flow and the turbulence level inside the room instead of only applying a correction based on the finite vertical extent of the upper and lower openings. These more complicated models are unfortunately time consuming to build and out of the boundaries of this project.

\subsection{Simulation of the external atmosphere} \label{sec:ideal-wind}
\subsubsection{Description of the problem}
\begin{figure}[ht]
\centering
  \includegraphics[width=\textwidth]{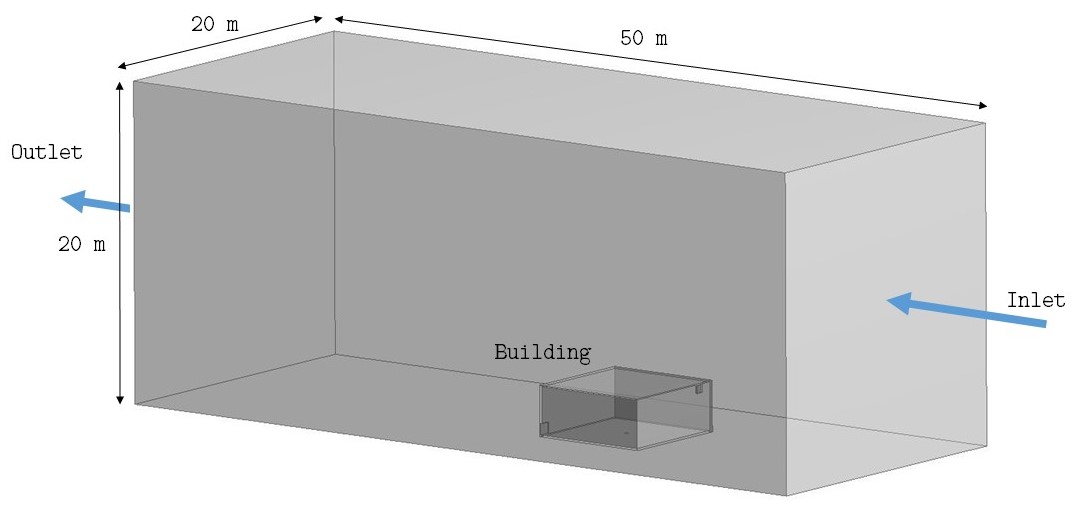}
  \caption{Illustration of the numerical domain used for the simulations with an ideal heat source and the effect of wind.}
  \label{fig:full_domain}   
\end{figure}

This set of simulations correspond to the theoretical model presented in section \ref{sec:external-atm-theo}.  Here wind is taken into account by including the external atmosphere in the numerical domain. The simulated external atmosphere was taken as a rectangle of 20m $\times$ 20m $\times$ 50m to avoid the influence of the wall boundary conditions on the internal flow (see Fig.~\ref{fig:full_domain}). The building geometry was taken from the previous (validated) numerical simulations.

\noindent The simulations were run for a building with a fixed normalized effective area of $A^*/H^2$ = 0.0210. This particular value was chosen deliberately because it was situated in the middle of the simulation range ($\xi \approx$ 0.5). Different wind speeds and their influence were studied. They are presented in Table \ref{tab:Flynn2009}.

\begin{table}[ht]
\centering 
\caption{Different cases simulated and their associated wind speeds}
\label{tab:Flynn2009}

  \begin{tabular}{l c c c c c c c c c c c}
  	
    Case & 1 & 2 & 3 & 4 & 5 & 6 & 7 & 8 & 9 & 10 & 11\\
   \hline
    Wind speed [m/s] & 0.1 & 0.2 & 0.25 & 0.3 & 0.35 & 0.4 & 0.5 & 0.6 & 1 & 1.5 & 2 \\

  \end{tabular}
\end{table}

\subsubsection{Mesh topology} \label{sec:ideal-wind-mesh}
\begin{figure}[ht]
\centering
\subfigure[Full mesh]{\includegraphics[width=0.4\textwidth]{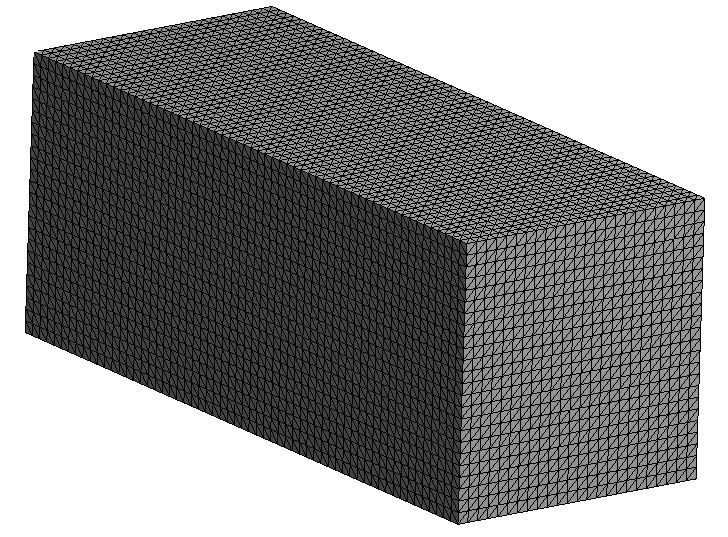}}
\subfigure[Details around the building]{\includegraphics[width=0.4\textwidth]{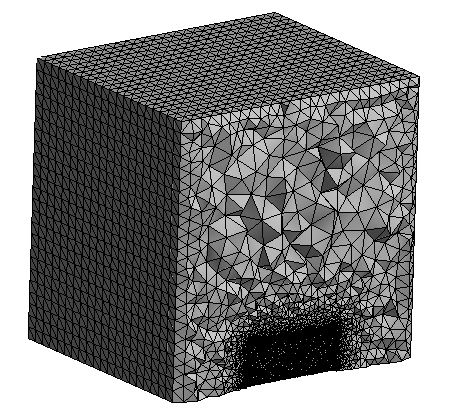}}           
\caption{Illustration of the mesh used for the simulations taking into account the influence of the external atmosphere.}
\label{fig:mesh-ideal-wind}
\end{figure}

The mesh used for the simulations taking into account the external atmosphere was based on tetrahedron cells. The reasons for this choice are the simplicity of generating such meshes since they provide easier meshing for complex geometries. To stay consistent with the convergence study performed for the building previously, the previous simulations where re-run with a tetrahedron mesh and proved to give the same results, though requiring a higher number of volumes. In this sense, the same mesh was used for the building inside the atmosphere. The tetrahedron-based mesh allowed the meshing of large cells in the simulated atmosphere whereas fine cells where used on the walls and inside the building, where finer flow phenomena occur. In this regard, if the use of this mesh works and succeeds in modeling the behavior of natural ventilation, it could be used in industry where results are needed quickly. Therefore, the mesh used is showed in Fig.~\ref{fig:mesh-ideal-wind}.

The important parameters assessing the quality of the mesh are regrouped in the table below (Table \ref{tab:mesh-external-atm}).

\begin{table}[ht]
\centering
\caption{Important properties assessing the quality of the mesh}
\label{tab:mesh-external-atm}

  \begin{tabular}{l c c c c c c}
   
    Nb.~elements & Max.~skewness & Min.~orth.~quality & Max.~aspect ratio\\
    1'000'000 & 0.23 & 0.76 & 1.86 \\
    
  \end{tabular}
\end{table}

\subsubsection{Operating conditions}
The operating pressure was set to 101325 Pa. For the full domain, since the pressure will directly enter in the validation of the model, the reference pressure for the numerical domain is probed 10 m upstream of the building, at ground level to avoid any thermal or wind influence. Density and temperatures are set to 1.225 kg/m$^3$, respectively 293 K.

\subsubsection{Boundary conditions}
The boundary conditions were defined as follow:
\begin{itemize}
	\item \textbf{\textit{inlet:}} velocity inlet with 0 Pa gauge pressure, and the speed specified in Table \ref{tab:Flynn2009}. Backflow temperature of $T_0$ and with very low level of turbulence ($k$ = 10$^{-5}$ and $\epsilon$ = 10$^{-6}$).
    \item \textbf{\textit{outlet:}} pressure outlet with 0 Pa gauge pressure, backflow temperature $T_0$ and very low level of turbulence ($k$ = 10$^{-5}$ and $\epsilon$ = 10$^{-6}$).
    \item \textbf{\textit{heat source:}} the heat source is modeled as a heating plate producing a heat flux corresponding to a power $P$ = 320 W.
    \item \textbf{\textit{building walls and floor:}} the walls and the floor are considered adiabatic with a standard no slip condition.
    \item \textbf{\textit{atmosphere walls:}} modeled as adiabatic walls with slip conditions to avoid the formation of boundary layers and reproduce the external atmosphere as well as possible.
	\item \textbf{\textit{ground:}} adiabatic wall with standard no slip condition.
\end{itemize}

\subsubsection{Discretization schemes}
All the terms were discretized with a second order scheme except for pressure where a PRESTO! scheme was chosen. The advantage of the latter is that instead of interpolating the pressure at the center of the cells, the pressure is directly computed on the faces. This scheme is more expensive computationally speaking but allows better precision on pressure, which is a key parameter for naturally ventilated flows.

\subsubsection{Wall treatment}
In the case studied in this project the precise modeling of the boundary layer only makes sense on the inside of the building where the walls are meshed more finely. The external atmosphere does not need to be discretized extensively since this would increase the required computational power to simulate a flow, but not give exploitable results. Therefore scalable wall functions are used. It allows the solver to choose whether a wall function or a near wall treatment is more appropriate. The use of this method is justified by the fact that velocity will vary along the simulations, along with the $Y^+$ value, it is therefore important that the solver gives solutions as close as possible to reality for all the steps of the simulation.

\subsubsection{Time-stepping}
The consideration of the external atmosphere in the numerical domain and the addition of wind to the simulations sets aside the possibility of using progressive time steps as it was done for the smaller domain (room only). The reason is that the addition of wind induces higher flow speeds inside the domain and for large time steps  numerical instabilities would be created in the simulated atmosphere, thus generating errors in the results. Due to this, the time step size is chosen constant according to the equation given in Fluent's user guide \cite{user_guide}:

\begin{equation} \label{equ:time_stepping}
\Delta t \leq \frac{L/4}{\sqrt{g \beta \Delta T L}} \approx \frac{L}{4 w}
\end{equation}

\noindent with L the characteristic length scale simulated ($L \sim$ 1m). The chosen time step is set to $\Delta t$ = 1 s in the case studied here, so that equ.~(\ref{equ:time_stepping}) is still valid for wind speeds up to 1 m/s. The time step has been lowered to $\Delta t$ = 0.5 s for higher wind speeds in order to avoid divergence and/or errors in the numerical results.

\subsubsection{Results} \label{sec:ideal-wind-results}
\textbf{Velocity field}\\
The different fields in the external atmosphere are not so important in the framework of this study, since the main interests are the physics happening inside the building especially because the useful value can be taken and calculated independently from the outside. Moreover, it would have been too time-consuming to verify the external parameters quantitatively. For these reasons only a qualitative analysis is done here, and the external velocity profile is plotted in Fig.~\ref{fig:external-atmosphere-streamlines} below. It appears that the profile is representative of a standard velocity profile that one could expect for the flow around a building. The streamlines show a disturbance of the flow above the building where the air is accelerated. On the other hand, a recirculation zone is created on the lee-side of the building. This is the origin of the pressure drop $\Delta p$ that will be used in the further calculations. \\

\begin{figure}[ht]
\centering
  \includegraphics[width=0.8\textwidth]{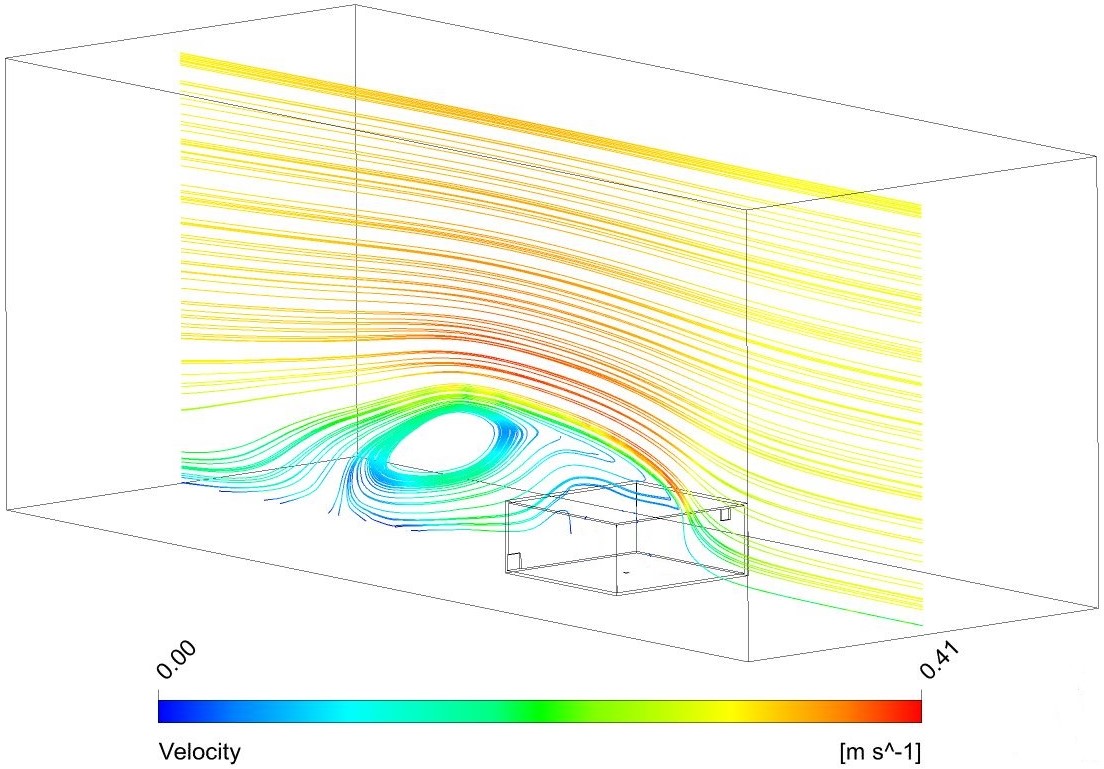}
  \caption{Streamlines of the flow in the middle of the domain for a inlet wind speed of 0.3\,m/s after 6000\,s.}
  \label{fig:external-atmosphere-streamlines}   
\end{figure}

\noindent In the case where a whole neighborhood was simulated, the domain would need to be bigger at least on the upper direction to avoid the over-acceleration on the flow above the building due to the squeezing of the area where the air flows. In any event, in the current case, it will not have an impact on the compared results since the pressure is taken independently from the upstream wind speed and the external velocity field.\\

\begin{figure}[ht]
\centering
  \includegraphics[width=\textwidth]{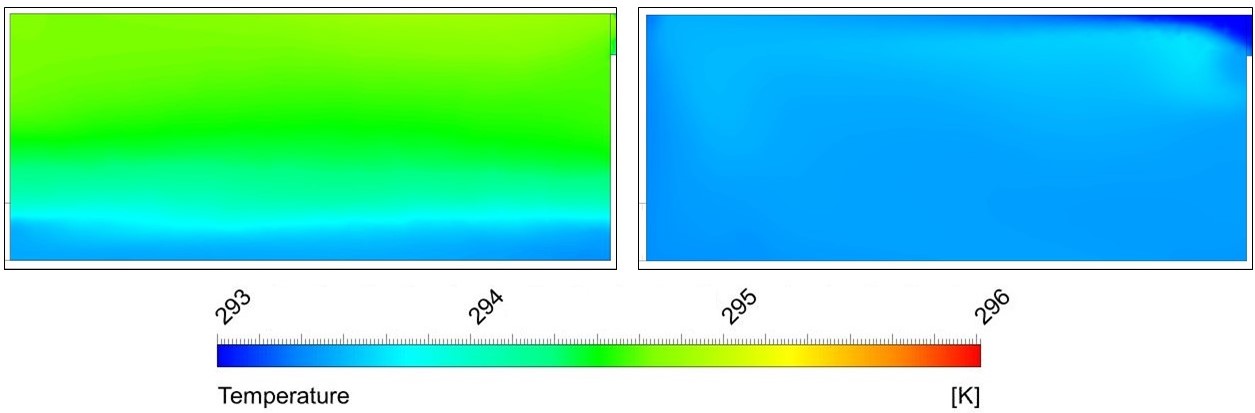}
  \caption{Temperature field inside the building for a wind speed of 0.3\,m/s (left) and a wind speed of 1\,m/s (right) after 6000\,s.}
  \label{fig:external-atmosphere-temp-field}   
\end{figure}

\textbf{Temperature field}\\
The temperature field at steady state inside the building is presented in Fig.~\ref{fig:external-atmosphere-temp-field} for a stratified case (left) and a mixed case (right).

\noindent For the stratified case, the density interface appears clearly on the figure, with a layer of cold air at $T_0$ and a warmer buoyant layer. It can be seen that the interface has a non-zero thickness, which puts in perspective the fact the data points compared with the theory will allow some incertitude \cite{Kayeflynn2010}. In the mixed case, the air inside the building is at an almost homogeneous temperature. The colder air in the top right corner corresponds to the cold external air entering the building.\\

\begin{figure}[ht]
\centering
  \includegraphics[width=0.8\textwidth]{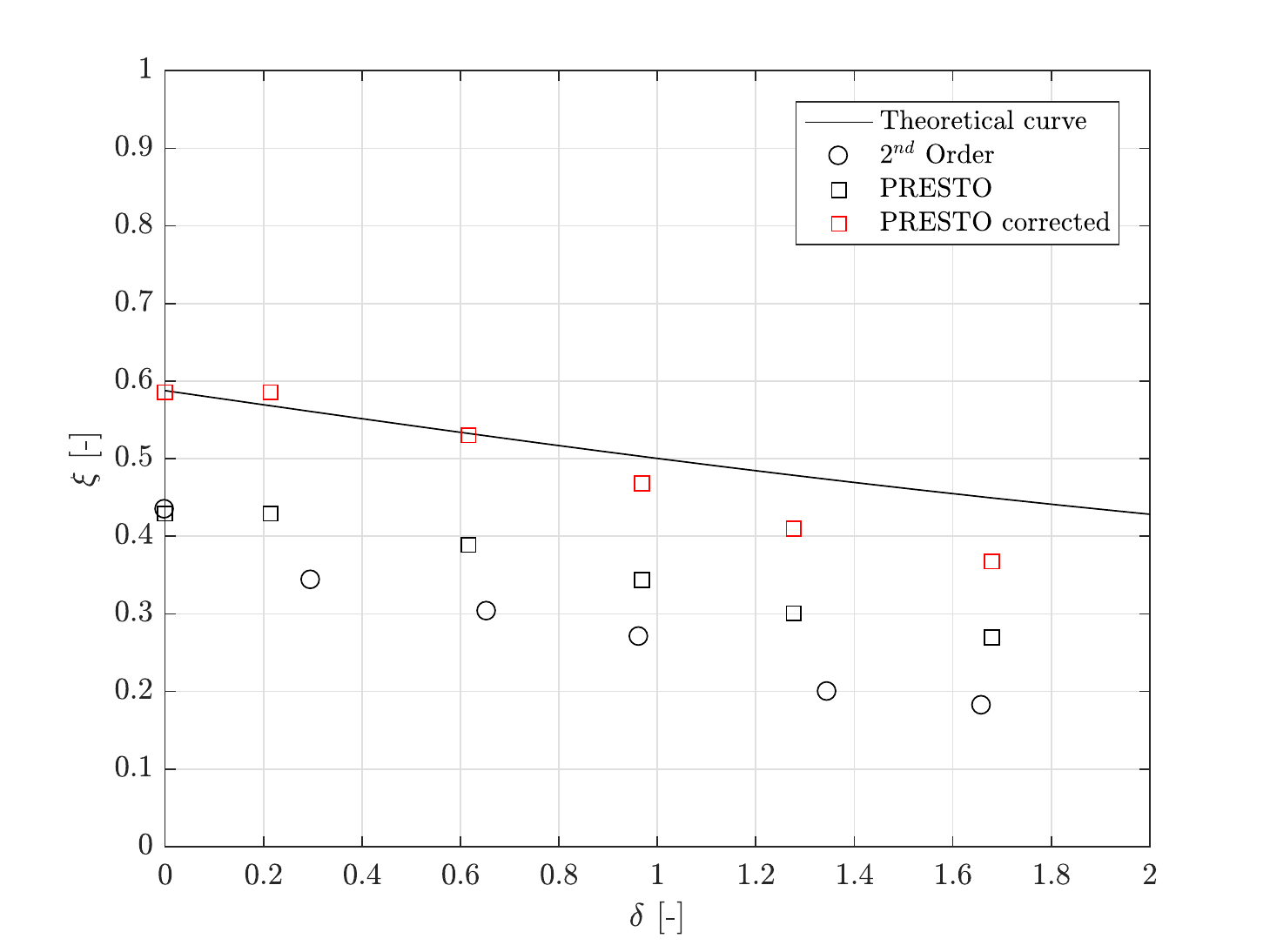}
  \caption{Non-dimensional interface height with respect to the non-dimensional parameter $\delta$ for a fixed normalized effective area of $A/H^2$ = 0.021. The different sets of curves are corresponding to different pressure discretization schemes. The red squares correspond to the data obtained with a corrected building height $H'$.}
\label{fig:xi_vs_delta_cfd}   
\end{figure}

\textbf{Interface height}\\
The numerical data obtained were added to the curves presented in the theory chapter (Fig.\ref{fig:xi_vs_delta}) for a normalized effective area of $A/H^2$ = 0.021 and are displayed in Fig.~\ref{fig:xi_vs_delta_cfd}. Of the first set of simulations, the pressure discretization was made with a standard second order discretization scheme and under-predicted the height of the density interface. Within Fluent's user guide \cite{user_guide}, it was recommended to use a PRESTO scheme along with the Boussinesq hypothesis. The simulations were re-run and the predictions were better, as it is shown in the figure below. Furthermore, the interface height was corrected using $H'$ instead of $H$ for the building interface height, accordingly to what was done previously. The final results seemed to agree fairly well with the theory.

\noindent It can be observed that the correlation between the numerical results and the theoretical model is diverging with the increasing wind speed (increasing $\delta$). This is likely due to some turbulent phenomenon that is not taken into account in the theoretical model. Moreover, it can be seen that the pressure discretization does not influence the case without wind ($\delta$ = 0). The reason is that this case does not take into account the external pressure drop in the calculations and is therefore not affected.

\noindent The data points not represented in Fig.~\ref{fig:xi_vs_delta} correspond to the mixed regime. Since no density interface is created, it would not make sense to add them on the graph. \\

\begin{figure}[ht]
\centering
  \includegraphics[width=0.8\textwidth]{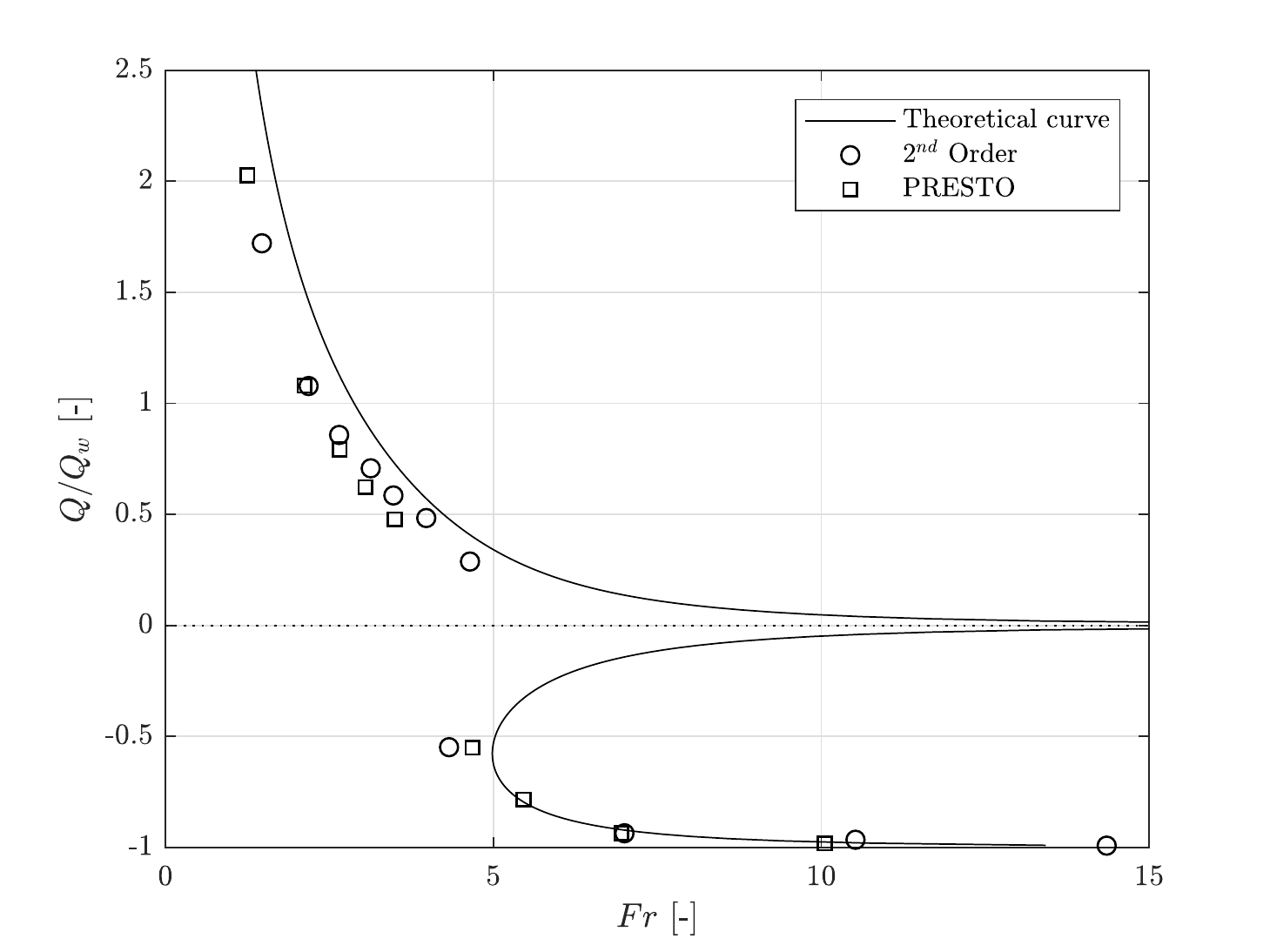}
  \caption{Variation of the normalized ventilation flux with respect to the Froude number.}
  \label{fig:QQw_vs_Fr_cfd}   
\end{figure}

\textbf{Ventilation flux}\\
The numerical results were added to the analytical curves presented in the theory (Fig.~\ref{fig:QQw_vs_Fr}) for a normalized effective area of $A/H^2$ = 0.021 and are displayed for both pressure discretization schemes (2$^{nd} $ order, PRESTO) in Fig.~\ref{fig:QQw_vs_Fr_cfd}. The figure shows good agreement for both schemes. It is interesting to note that since the simulations are initialized with a zero speed everywhere, the solution will never fall to the right of the critical Froude number, defined as that minimum value of Fr where the well-mixed solutions first appear. The numerical model is also performing well at predicting the state of the flow (stratified or mixed).

\noindent This validates the numerical model for the simulation of the building and the external atmosphere. Since the ventilation flux is a parameter that is not directly linked to the effective height of the building, it does not need to be corrected with the same procedure than the interface height.

\subsection{Simulation of the blocked ventilation}
\subsubsection{Description of the problem}
For this set of simulations, the effect of a non-ideal heat source are studied but in the absence of an external wind or, for that matter, an explicitly modeled external atmosphere. To this end, the numerical domain of the first set of simulations is used since the only difference is that this time the source provides mass to the system. Therefore, the boundary conditions are the same as before, as well as the discretization schemes and the mesh. The different volume source flux and the associated $\zeta_s$ values are summarized in the table below.

\begin{table}[ht]
\centering
\caption{Different volume sources simulated and the associated non-dimensional source flux values for a normalized effective area of $A^*/H^2$ = 0.021.}
\label{tab:non-ideal-cases}

\begin{tabular}{l c c c c c c c }
	 Case & 1 & 2 & 3 & 4 & 5 & 6 \\
	$Q_s$ [m$^3$/s] & 0.04 & 0.06 & 0.08 & 0.10 & 0.12 & 0.14   \\
    
	$\zeta_s$ [-] & 0.41 & 0.48 & 0.53 & 0.58 & 0.63 & 0.67  \\
       
\end{tabular}
\end{table}

\subsubsection{Boundary conditions}
The boundary conditions are the same as the ones presented in section \ref{sec:bc-ideal-no-wind}. The only change is the heat source, which now provides mass to the system:
\begin{itemize}
	\item \textbf{\textit{heat source:}} the heat source is modeled as a mass flow inlet providing air at 300 K and for the values presented in Table \ref{tab:non-ideal-cases}
\end{itemize}

\subsubsection{Results} \label{sec:non-ideal-no-wind-results}
Since the qualitative results were exhibiting the same tendencies as the ones presented in the section before (section \ref{sec:ideal-wind-results}), they are not detailed here. Only the quantitative results compared with the theoretical model will be discussed.\\

\begin{figure}[h]
\centering
  \includegraphics[width=0.8\textwidth]{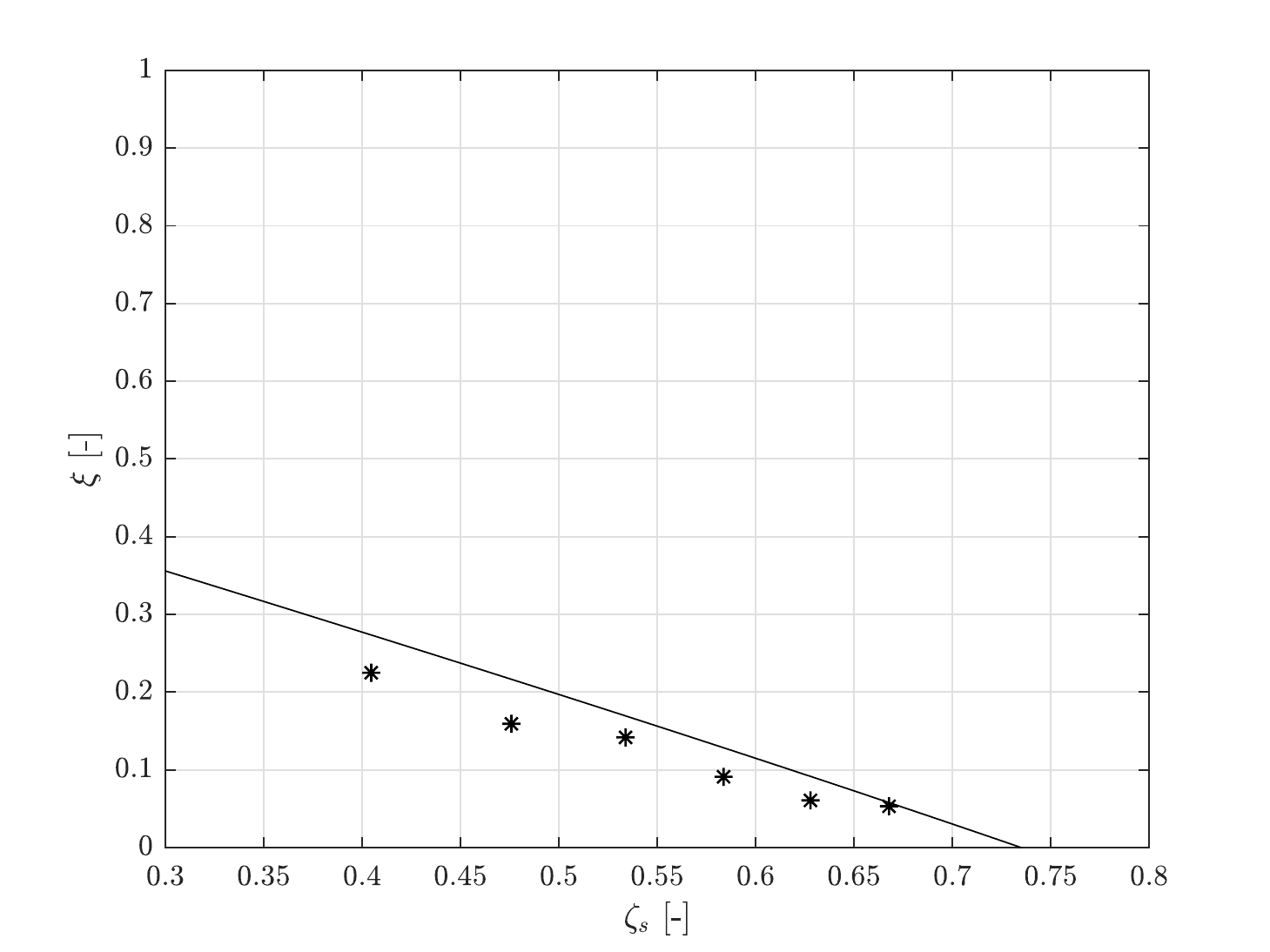}
  \caption{Variation of the non-dimensional interface height with respect to the non-dimensional volume heat source for a fixed normalized effective area of $A^*/H^2$ = 0.021.}
  \label{fig:xi_vs_zetas_cfd}   
\end{figure}

\textbf{Interface height}\\
Fig.~\ref{fig:xi_vs_zetas_cfd} represents the non-dimensional interface height with respect to the non-dimensional volume heat source. The black line is the theoretical line presented in section \ref{sec:non-ideal-no-wind} and the black dots correspond to the different numerical results.

\noindent The results are fitting fairly well to the theory. In general, the interface height is slightly under-predicted by the numerics but the tendency is correct. Since the simulations have been run with the top/bottom openings on the ceiling, respectively on the floor of the room, the results do not need to be corrected with $H'$ instead of $H$. 

\subsection{Simulation for a non-ideal heat source} \label{sec:non-ideal-wind}
\subsubsection{Description of the problem}
The simulations carried out in this section are corresponding to the theory depicted in section \ref{sec:therory-full-non-ideal}. The numerical domain is the same that was used for the first set of simulations taking into account the influence of the external atmosphere, as well as the mesh (see section \ref{sec:ideal-wind-mesh}). The boundary conditions are kept the same except for the heat source that now provides mass as well as buoyancy The discretization schemes are of the second order, except for pressure for which the influence of a PRESTO discretization scheme was studied. Two different non-ideal heat sources ($\zeta_s = \{0.23, 0.58\}$) were tested, as well as nine different wind speeds. They are summarized in Table \ref{tab:flynn2009-MF}.

\begin{table}[ht]
\centering
\caption{Different wind speeds studied for the simulations of the non-ideal heat source and the influence of the external atmosphere.}
\label{tab:flynn2009-MF}

\begin{tabular}{l c c c c c c c c c c}
    
	 Case & 1 & 2 & 3 & 4 & 5 & 6 & 7 & 8 & 9 \\
    \hline
	Wind speed [m/s] & 0.1 & 0.2 & 0.25 & 0.3 & 0.35 & 0.4 & 0.5 & 0.7 & 1\\
       
\end{tabular}
\end{table}

\subsubsection{Results}
Since the qualitative results such as the temperature and velocity fields are exhibiting the same behavior than what was presented in section \ref{sec:ideal-wind-results} they are not detailed here. The discussion will rather focus on the quantitative aspect of the results.\\

\begin{figure}[ht]
\centering
  \includegraphics[width=0.8\textwidth]{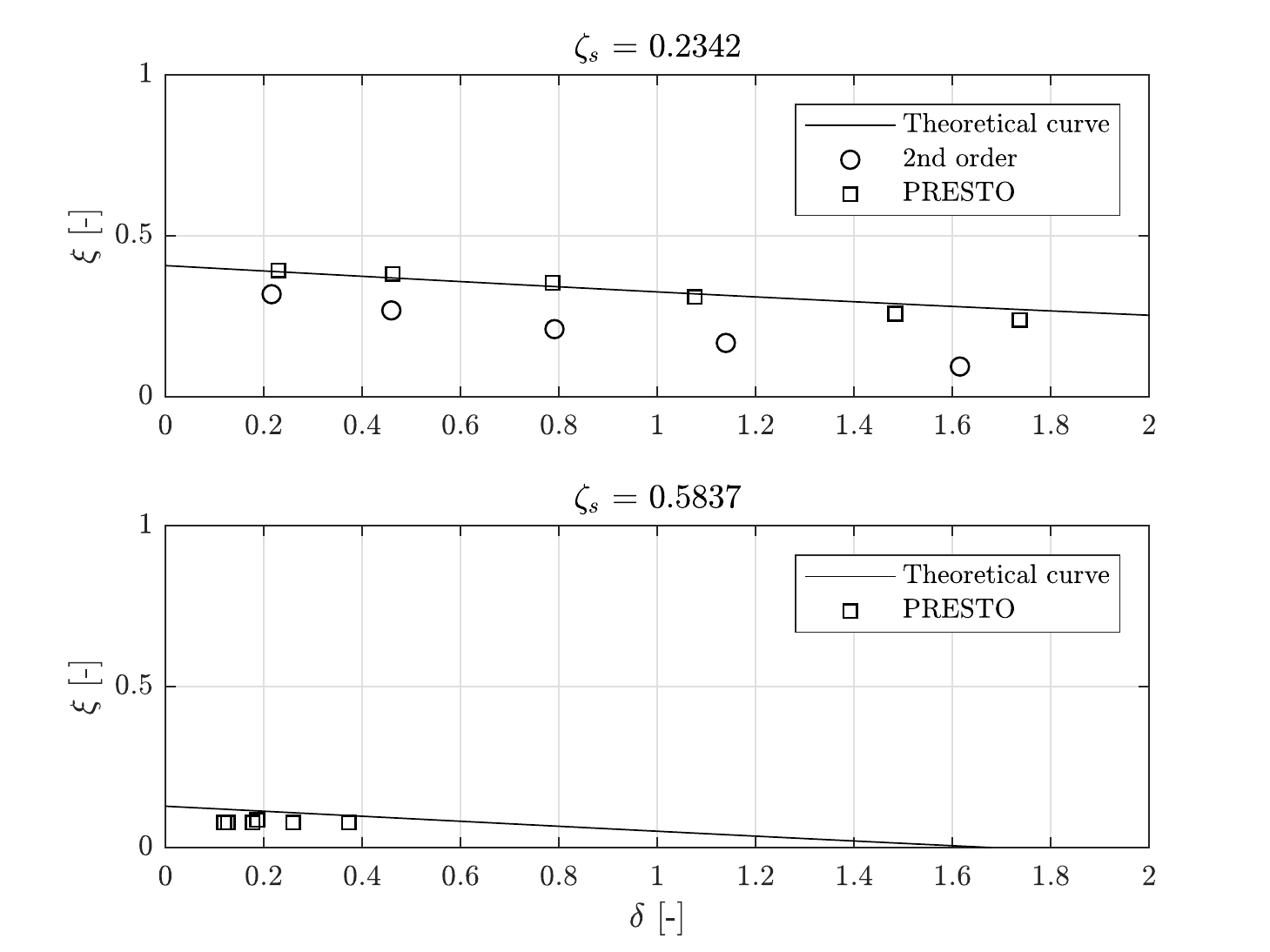}
  \caption{Variation of the non-dimensional interface height with respect to the non-dimensional parameter $\delta$ for a fixed normalized effective area of $A^*/H^2$ = 0.021.}
  \label{fig:xi_vs_delta_MF_cfd}   
\end{figure}

\textbf{Interface height}\\
Fig.~\ref{fig:xi_vs_delta_MF_cfd} represents the non-dimensional interface height with respect to the non-dimensional parameter $\delta$. The black line corresponds to the analytical curve, the black circles to the numerical results for a second order pressure discretization scheme and the black squares to the numerical results for a PRESTO pressure discretization scheme.

\noindent The influence of the discretization scheme for pressure is shown. The PRESTO scheme appears to perform better at predicting the interface height. It means that the pressure is a key parameter that influences the location of the buoyant layer in naturally ventilated flows. The upper graph in Fig.~\ref{fig:xi_vs_delta_MF_cfd} represents an average case for the volumetric heat source and theory and numerics agree fairly well with theory. On the other hand, for strong heat source volumes fluxes, such as the lower graph in the figure, the interface height is lowered till the upper boundary of the bottom opening due to the forcing of the source. As a results the interface goes down to its physical limit and not lower but because numerically (and physically) the interface height cannot be lowered under the upper boundary of the bottom opening. As the density interface lowers it will reach the bottom opening at one point and the the air from the buoyant layer will escape through that opening as well. Consequently, the second set of simulations cannot serve as a validation of the theory, even though the results seem to be in general agreement.\\

\begin{figure}[ht]
\centering
  \includegraphics[width=0.8\textwidth]{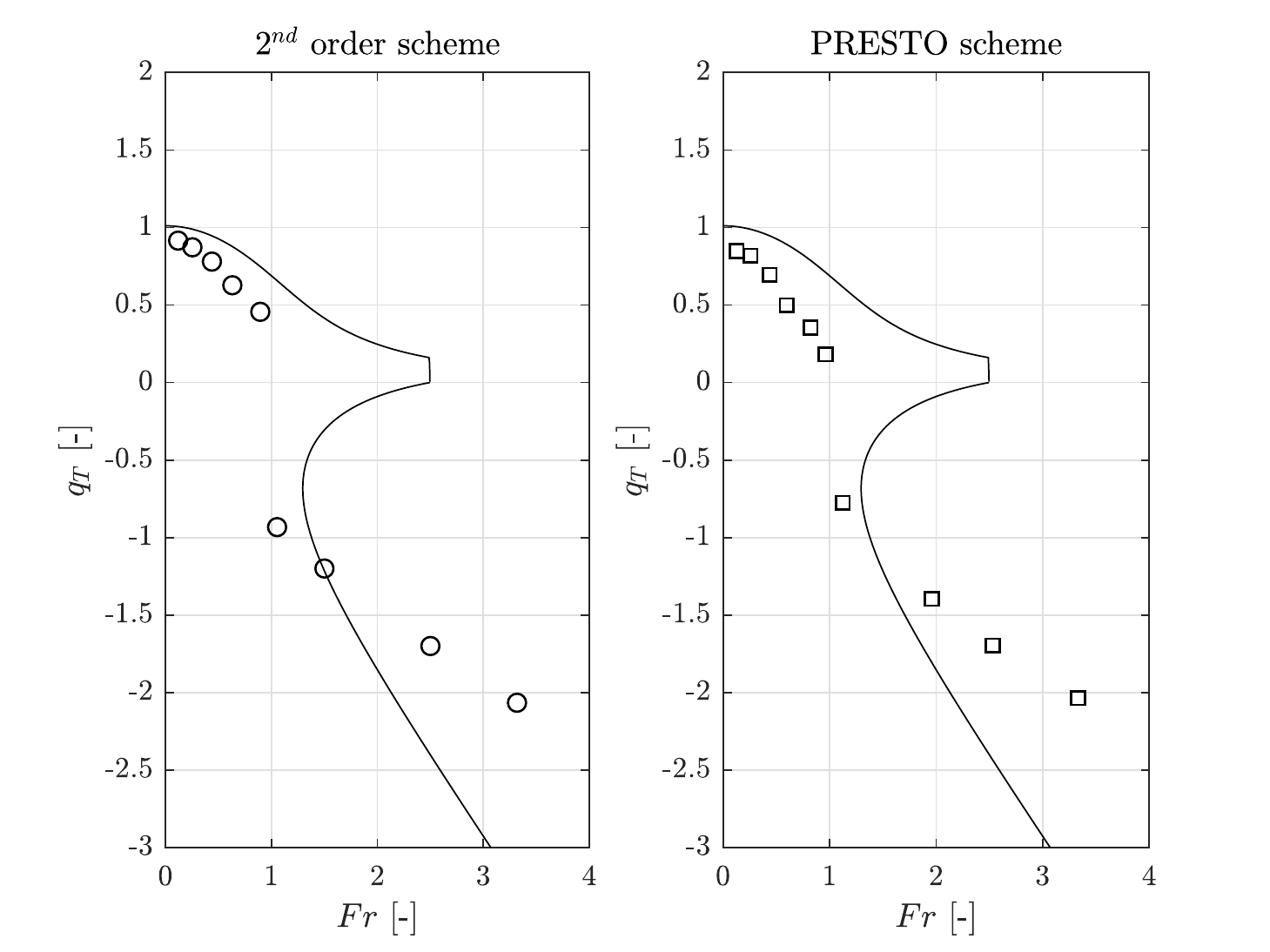}
  \caption{Variation of the normalized flow rate with respect to the Froude number for a non-ideal heat source and a fixed normalized effective area of $A^*/H^2$ = 0.021.}
  \label{fig:qT_vs_Fr_cfd}   
\end{figure}

\textbf{Flow rate}\\
Fig.~\ref{fig:qT_vs_Fr_cfd} represents the normalized flow rate in function of the Froude number, corresponding to the Fig.~\ref{fig:qT_vs_Fr} presented in the theory.

\noindent The pressure discretization scheme does not have an influence on the behavior of the flow rate, both curves follow the same tendency. The numerics are failing to properly model the flow when the state should be blocked. Instead, the same discontinuous behavior that was observed for an ideal source is seen, with a flow rate that jumps from positive to negative values when close to zero. More than that, the tendencies are not fitted correctly.\\

Regarding the fact that the lower part of the curve (mixed state) does not follow the good tendency, it would not be appropriate to draw conclusions from this figure. Further studies should focus on fitting the curve before trying to understand why the behavior predicted by the analytical model does not agree with the numerical simulations. In parallel, a experimental campaign should be carried out since the numerical model has not been validated yet. It is therefore not possible to say at this point whether the theory or the numerics (or both) are failing.

\subsection{Simulation of the decoupled heat source}
\subsubsection{Description of the problem}
\begin{figure}[ht]
\centering
  \includegraphics[width=\textwidth]{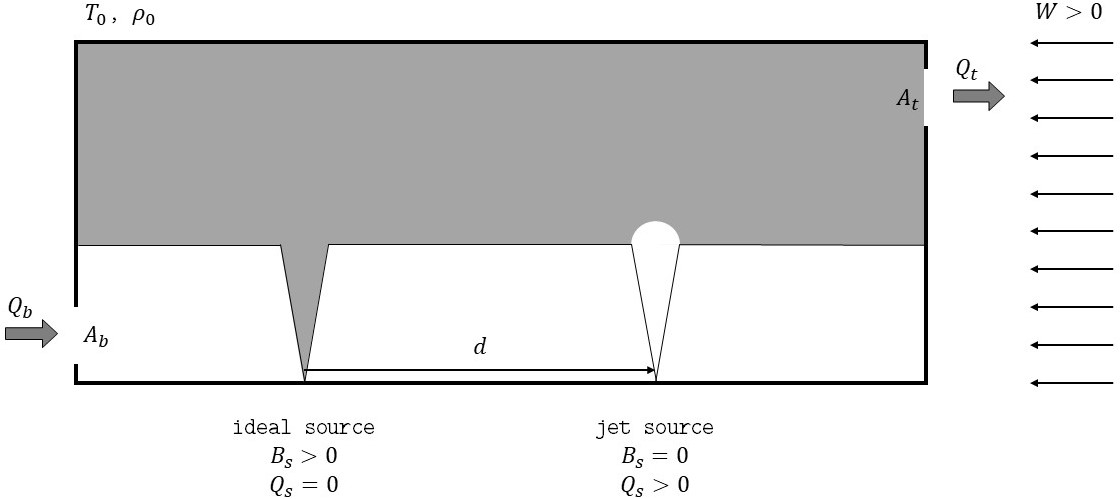}
  \caption{Illustration of the idealized considered geometry and the decoupled heat source.}
  \label{fig:slide4}   
\end{figure}

The geometry presented in Fig.\ref{fig:one_box_building} and the related atmospheric parameters are kept the same as before. Instead of having a single heat source in the center of the room, the source is decoupled. The first source (represented on the left in Fig.~\ref{fig:slide4}) will provide only heat to the internal atmosphere, whereas the second source (represented on the right in Fig.~\ref{fig:slide4}) will provide a source volume flux at $T_0$ (air jet).

\noindent To asses the effects of the decoupling, the influence of the distance $d$ between the two sources is studied. $d$ is the distance of the ideal source on the left of the jet, so that if the sources are inverted and the ideal source is on the right, $d$ becomes negative. To assess the influence of $d$, different wind speeds, are simulated for $d = \{-H, -\frac{H}{2}, \frac{H}{2}, H \}$. The mesh used is the same that was used for the first set of numerical simulations (see section \ref{sec:ideal-no-wind-mesh}) as well as the discretization schemes (PRESTO for pressure) and the boundary conditions.

\subsubsection{Results}
As mentioned no theoretical model nor experiments were made for decoupled sources. The advantage of studying numerically the decoupling of the heat source is that it provides a fast and relatively reliable way of assessing if decoupling has an influence on the flow inside the building.\\

\begin{figure}[ht]
\centering
  \includegraphics[width=\textwidth]{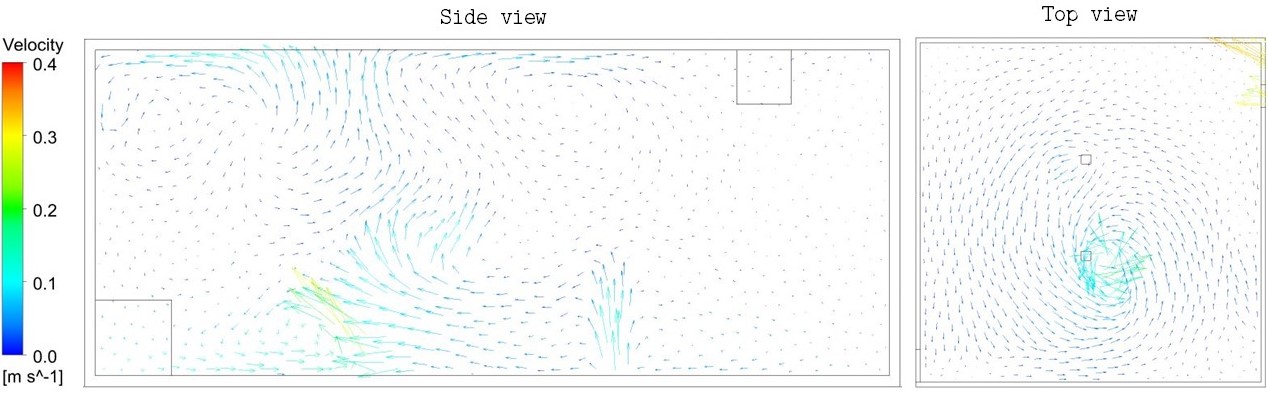}
  \caption{Velocity field inside the simulated building for a fixed effective area of $A^*/H^2$ = 0.021 and an ideal source of 80 W and a jet of 0.01 m$^3$/s. The two source are separated by a distance of $d$ = $H/2$.}
  \label{fig:decoupledHs-vel-field}   
\end{figure}

\textbf{Velocity field}\\
Fig.~\ref{fig:decoupledHs-vel-field} represents the velocity field inside the building for an ideal source of 80 W, an air jet of 0.01 m$^3$/s and a fixed normalized effective area of $A^*/H^2$ = 0.021. The left part of the figure shows the velocity field on a plane parallel to the front wall (wind-side) of the building, at the center and the right part the velocity field plotted on the density interface.

\noindent The velocity field clearly shows that the flow is dominated by buoyancy and not by the air jet. This is mainly seen on the top view where high velocities are composing the field around the heat source and not the jet. The turbulence level created by the heat source is also significantly superior to the one coming from the air jet. As a result, the velocity field inside the building for a decoupled source is very similar to that observed in the case of a single source, for both ideal and non-ideal cases.\\

\begin{figure}[ht]
\centering
  \includegraphics[width=\textwidth]{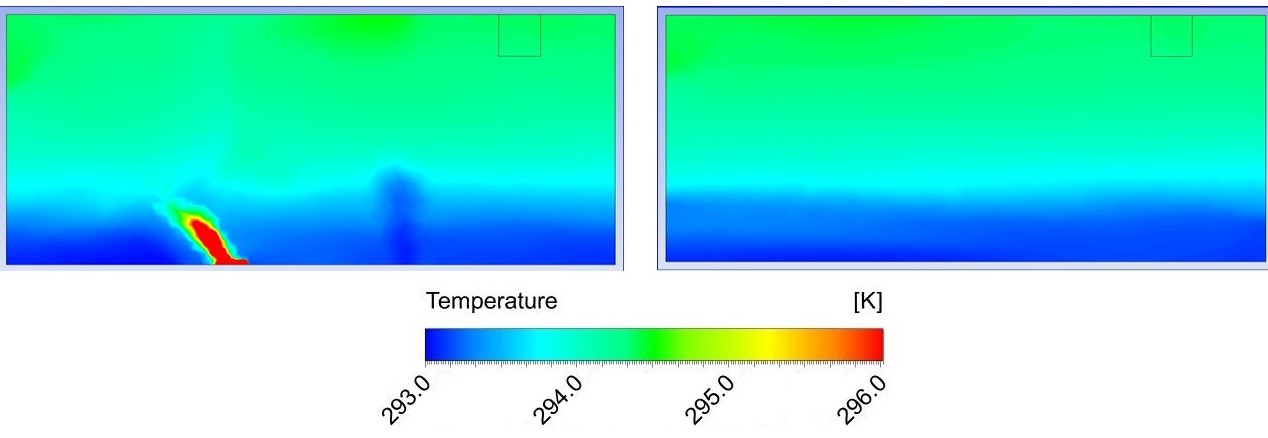}
  \caption{Temperature field inside the simulated building for a fixed effective area of $A^*/H^2$ = 0.021 and an ideal source of 80 W and a jet of 0.01\,m$^3$/s. The two source are separated by a distance of $d$ = $H/2$.}
  \label{fig:decoupledHs-temp-field}   
\end{figure}

\textbf{Temperature field}\\
Fig.~\ref{fig:decoupledHs-temp-field} represents the temperature field inside the building for the same parameters presented before on a plane parallel to the front wall at the center of the sources and at 1.5 m off the center.

\noindent The two sources can be distinguishably seen on the figure above. The fact that the interface height is well defined above the air jet but more heavily distorted above the heat source confirms the observations made with the velocity field. For the parameters simulated, the buoyancy is dominating the flow. Moreover, the air provided by the air jet seems to stay in the lower density layer and is just creating a bump on the interface by pushing upwards, locally.\\

\begin{figure}[ht]
\centering
  \includegraphics[width=0.8\textwidth]{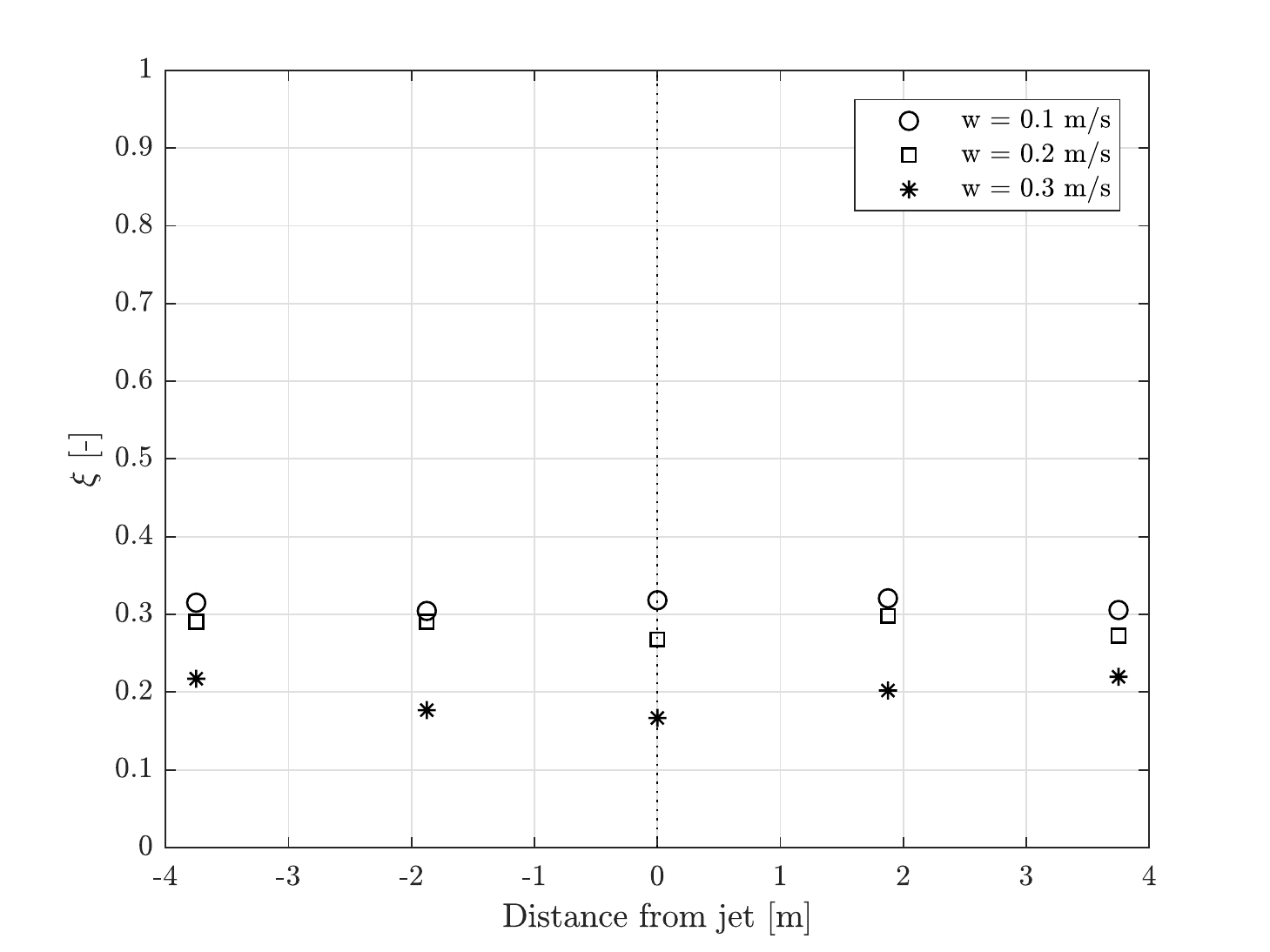}
  \caption{Variation of the non-dimensional interface height with respect to the distance $d$ between the sources for different wind speeds.}
  \label{fig:xi_vs_d}   
\end{figure}

\textbf{Interface height}\\
Fig.~\ref{fig:xi_vs_d} represents the non-dimensional interface height with respect to the distance $d$ for different wind speeds (and therefore different $\delta$). The data point at $d$ = 0 m are corresponding to a ``coupled'' source, i.e. a source that provides both heat and mass to the system with a corresponding source strength given by $\zeta_s$ = 0.2343. Concretely, these data points were added from the simulations run for a non-ideal heat source with the influence of the external atmosphere (section \ref{sec:non-ideal-wind}). To insure the validity of the comparison, an equivalent buoyancy flux was chosen for the source ($B$ = $B_s$).

\noindent It appears that the distance $d$ does not influence the density interface height. The differences that can be seen on the figure above are likely due to the turbulent nature of the flow inside the domain and not a relevant discrepancy. To support this argument, it can be seen that the results are more stable for low wind speed, thus lower turbulence levels. Concretely, it means that it is unnecessary to conduct mathematical investigations where the source is decoupled at least while trying to improve the prediction of the interface height.\\

The impact of the flow rate was not studied in the results since the previous results for a non-ideal heat source showed that the  tendencies were not correctly modeled by the numerics. It is therefore not possible to extend the results and draw conclusions.


\section{Discussion and perspectives}

\subsection{Correction model for the non-dimensional density interface height}
The first thing that needs to be pointed out is the way the correction model for the density interface height prediction is built. Since the simulations with a non ideal heat source are fitting well the theory without applying a correction to the non-dimensional interface height emphasizes the fact that using a shorter building height to correct the results might not be the optimal way to do it. As it was already mentioned in the presentation of the results, a corrective model taking into account the inner vorticity of the flow and/or the turbulence level of the domain would certainly have led to a better fit of the model. Moreover, the model should also take into account the type of source since it seems to have an impact on the final prediction.\\

\begin{figure}[ht]
\centering
  \includegraphics[width=0.8\textwidth]{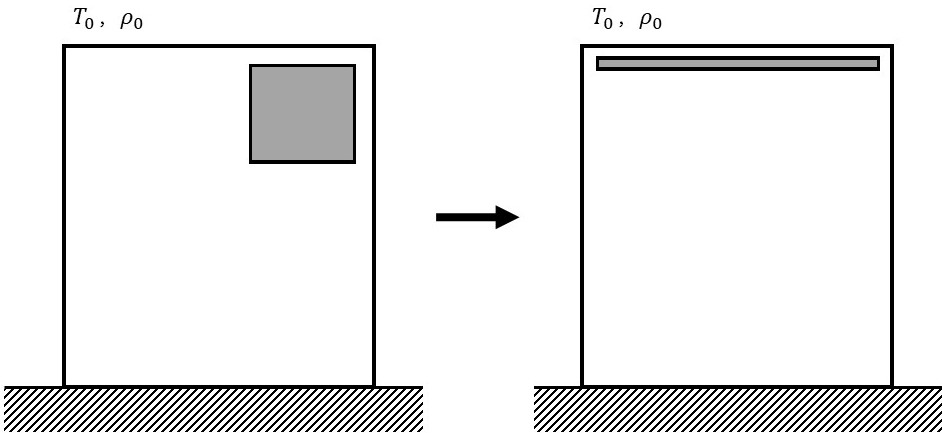}
  \caption{Different geometries of lateral openings for the same normalized effective area.}
  \label{fig:different_effective_area}   
\end{figure}

To avoid the long process involved, an option would be to modify the type of openings used in the simulations. Even though square openings were easier to model and facilitated the comparisons, they are not optimal since, as explained above it might have an impact on the ``real'' height of the building. Considering different geometries for the openings for instance as presented in the figure below (Fig.~\ref{fig:different_effective_area}) could help with the problem.\\

Moreover, the models described above are trying to represent real buildings where the openings are all the thin cracks and small openings that allow a building to ``breathe,'' not only the build pipes that allow ventilation to happen. By doing so, one must be aware that changing the geometry of the openings will have an impact on the loss coefficient $c_b$ and $c_t$ that will no longer be approximated by the 0.6 value. Many studies have nevertheless been conducted to study their variations and good papers treating this subject can be found (e.g. \cite{Karava2004,Chia2009}). Finally to validate those coefficients, simulations could be run with an emptying box, and the discharge coefficient assessed from the emptying time of the room.

\noindent In this context, one could think that measuring the interface height is maybe not the optimal parameter to assess the correct modeling of the flow. The fact is that using the height of the density interface as a metric to determine the validity of a model comes from the experimental campaigns. In real-life experiments, all the data such as velocity, temperature, pressure fields, that can be relatively easily known via analysis or numerics are extremely hard to measure precisely. As an example, imagine how hard it is to measure precisely the pressure drop between the top and bottom openings of the building without interfering with the flow. In this sense, the measurement of the density interface height can be made relatively quickly (and not too costly) with optical methods.

\subsection{Simulation of the external atmosphere}
The simulation of the external atmosphere was here done directly in the sense that the whole domain was simulated. The main reason for that was to be as close to reality as possible. For instance, putting a constant pressure drop over the building might have led to results with a better agreement with theory but further from reality. In this sense to improve the efficiency of the simulations without going too far from reality, one option could have been to run simulations of a wind flow around a closed building, for the different studied wind speeds and save the velocity (temperature and pressure) field as a User Defined Function. In a second time, another set of simulations would have been run, using the field computed before as inlets, this under the assumption that the pressure field is comparable between the closed and ventilated building. The problem is that it is hard to assess the numerical behavior of the interaction on the boundaries (e.g. when the air inside the building starts escaping the room through the top opening) if an unsteady boundary condition is fixed. For this reason mainly, the simulations were run as is.

\subsection{Decoupling of the heat source}
Specifically concerning the decoupling of the source, the relative importance of the heat source vs. the air jet should be studied. The simulations run above were clearly dominated by the heat source and therefore might have behaved differently if the air jet was stronger. Also, the inlets were specified to have very low levels of turbulence, but as it was mentioned a few times, some discrepancies between the results have certainly turbulence as origin, and it might be interesting to investigate in this direction.



\section{Conclusion}

During this project, the feasibility of simulating naturally ventilated flows in buildings was assessed. The goal was not only to try to model these flows with numerical simulations but to show that all the different regimes that take place can be reproduced with computational fluid dynamics. To do so, existing analytical models were regrouped and standardized, particularly through the reorganization and re-writing of the existing equations incorporating the same parameters. Besides, numerical simulations corresponding to the different analytical models were run and then compared. The simulations showed good agreement with the theory for most cases. Since some of the theoretical work was previously validated, the results can be used to validate the numerical solutions and their ability to predict natural ventilation flows. In addition, the influence of a decoupled heat source was studied. The main objective was to assess if a decoupled heat source could improve the prediction of the flow parameters. The results show that at least for the prediction of the density interface height, this was a priori not the most promising way to investigate. With that being said, one must keep in mind that complementary simulations should be run before drawing a final conclusion.\\



\newpage
\bibliographystyle{model1-num-names}
\bibliography{refs.bib}







\newpage
\appendix
\section{Measurement of the numerical data} \label{app:measurements}

\subsection{Pressure drop over the building and volumetric fluxes through the openings}
The pressure drop over the building must be known to compute the non-dimensional parameter $\delta$. The pressure values are averaged over the top and bottom opening areas respectively. Moreover, since the values are slightly fluctuating due to turbulence, a time-average is taken from the moment the flow is well established inside the domain (the pressure has reached a steady-state value). A similar method was applied for the computation of the volumetric fluxes flowing in and out of the building.

\begin{figure}[ht]
\centering
  \includegraphics[width=0.65\textwidth]{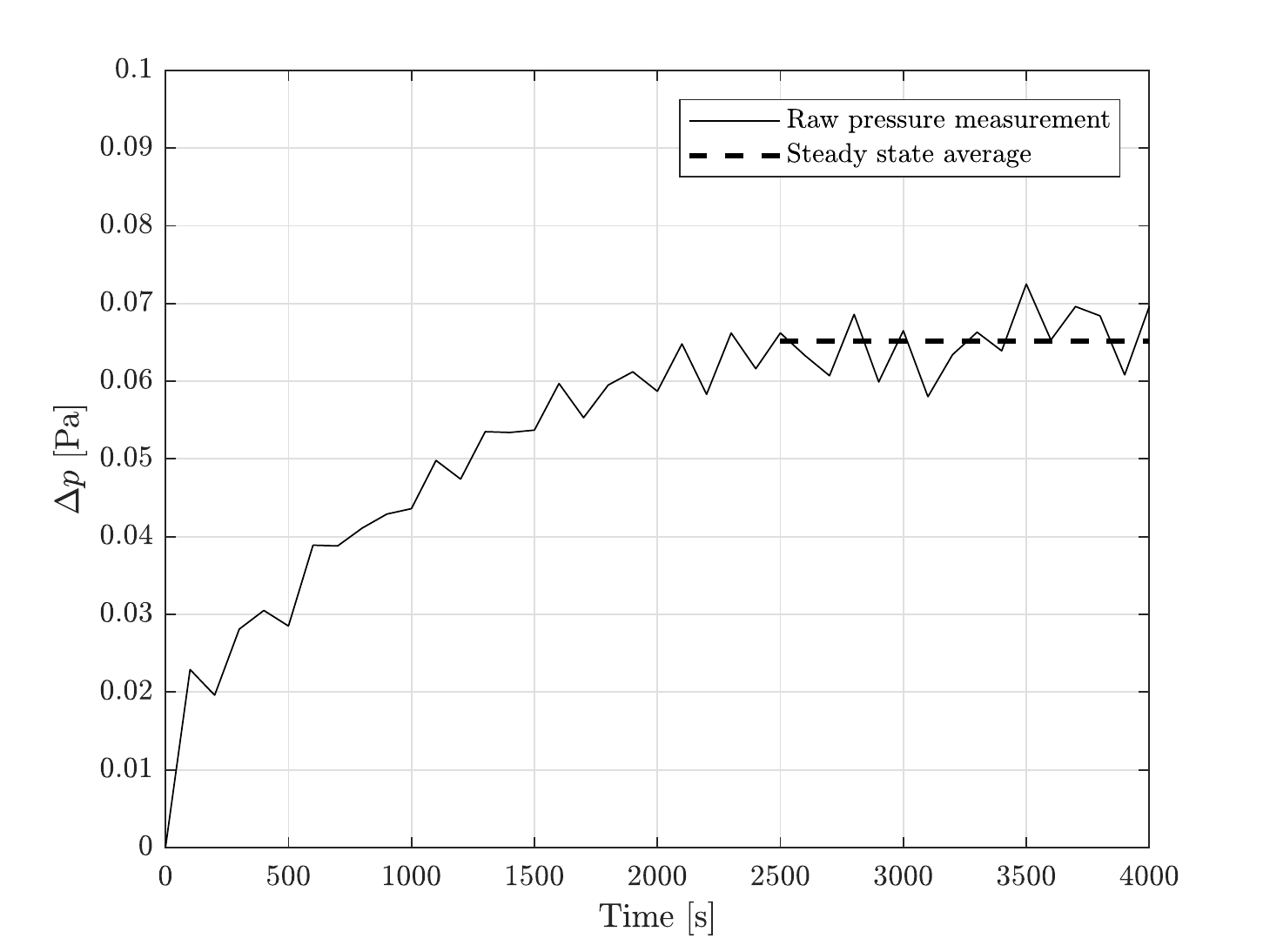}
  \caption{Evolution of the pressure drop $\Delta p$ over the wind and lee-sides of the building. The dashed line represents the averaged value when steady state has been reached.}
  \label{fig:deltap_vs_time}   
\end{figure}

\subsection{Interface height}
\begin{figure}[ht]
\centering
  \includegraphics[width=\textwidth]{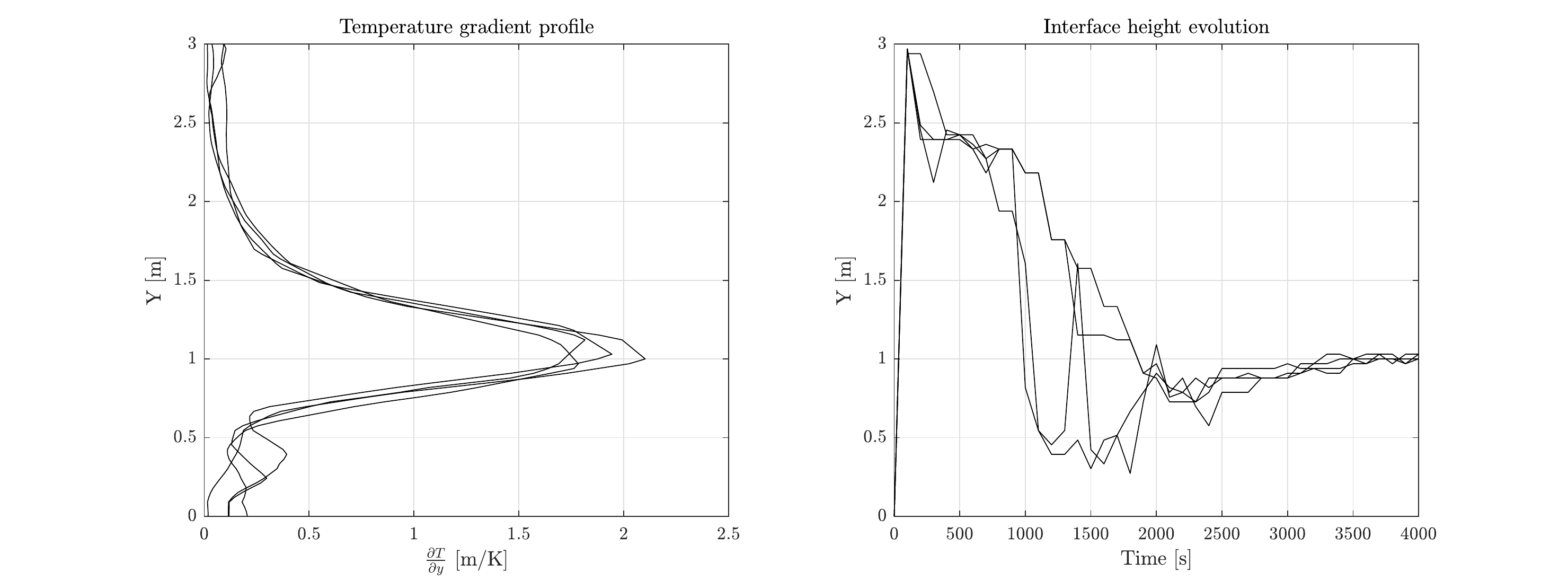}
  \caption{Temperature gradient profile over the height of the building (left) and evolution of the density interface height over time (right).}
  \label{fig:intheight_vs_time}   
\end{figure}

Four probes are created inside the building, at different horizontal locations where turbulence coming from the openings or the floor source(s) have low influence. The temperature gradient (Fig.~\ref{fig:deltap_vs_time}) is computed over those lines and the maximum value is taken as the location of the density interface. Generally, the maximums have converged to the same value, if this is not the case, the mean height of the four probes is taken as the interface height. From there, the evolution of the interface height can be determined over time (Fig.~\ref{fig:intheight_vs_time}).\\

\noindent The right part of the figure shows that during the establishment of the steady state flow, the turbulence level inside the building is high and stabilizes with time.\\

Another promising approach would be to numerically create planes parallel to the ground, spaced by a few centimeters and average the temperature on them instead of using the technique mentioned above. Like that, the temperature (or other variables) field would have been reduced to one direction only and the results might be closer to the theory. The main reason why this technique was not used is the huge amount of computational resources that is required to average one variable on many different planes and over long numerical periods of time.

\subsection{Temperature of the thermal layers}
\begin{figure}[ht]
\centering
  \includegraphics[width=0.8\textwidth]{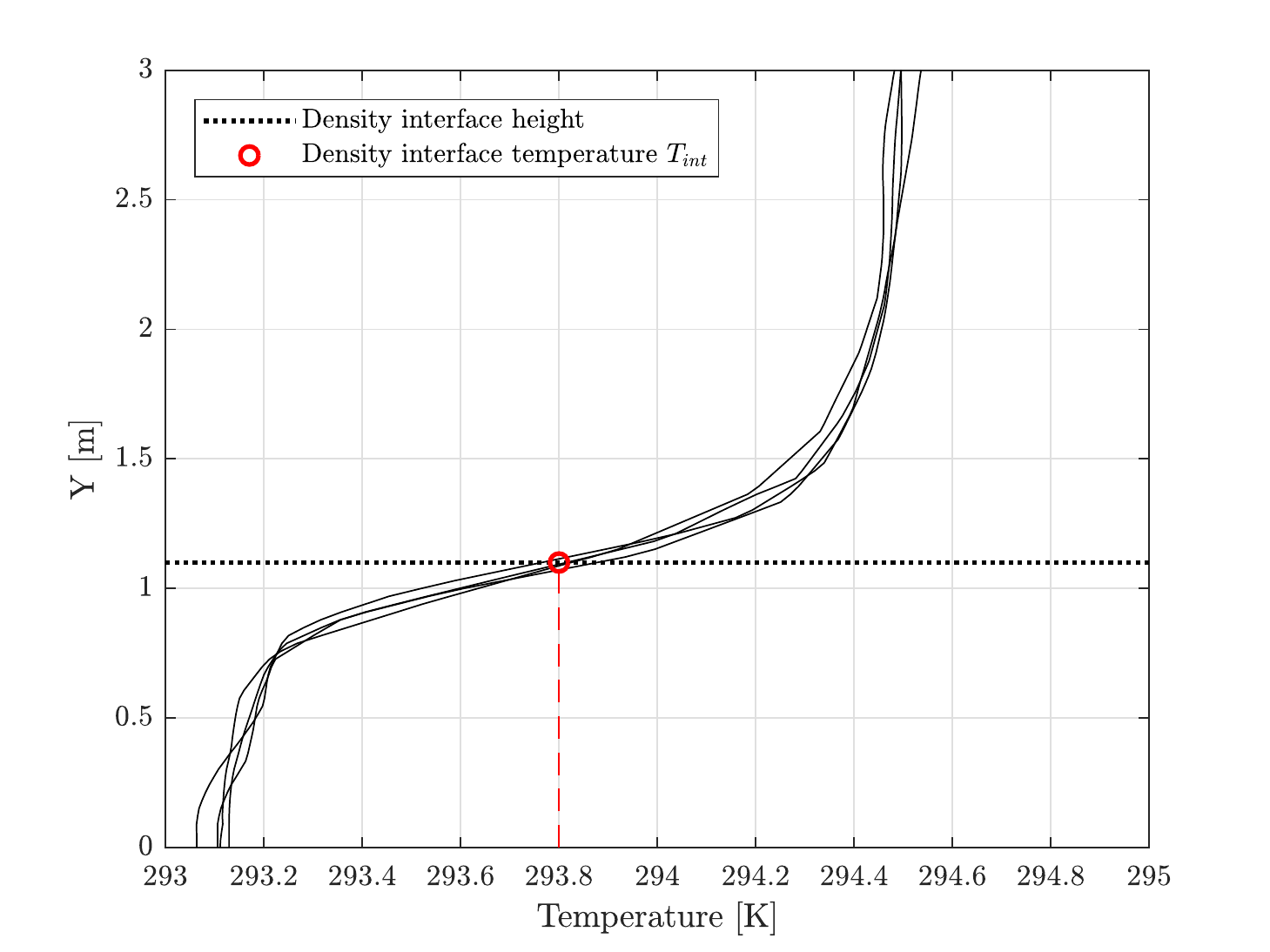}
  \caption{Temperature profile of inside the building for the four different probes. The dashed line represents the density interface and the red dot the corresponding temperature.}
  \label{fig:y_vs_temp}  
\end{figure}

As explained before, the density interface height can be established from the temperature gradient profile inside the room. Since the temperature profile can also be established, it is possible to compute the temperature $T_{int}$ of the density interface, as shown on Fig.~\ref{fig:y_vs_temp}. 

\begin{figure}[ht]
\centering
\subfigure[Numerical surface at the temperature of the density interface height $T_{int}$.]{\label{fig:isosurf_w025} \includegraphics[width=0.4\textwidth]{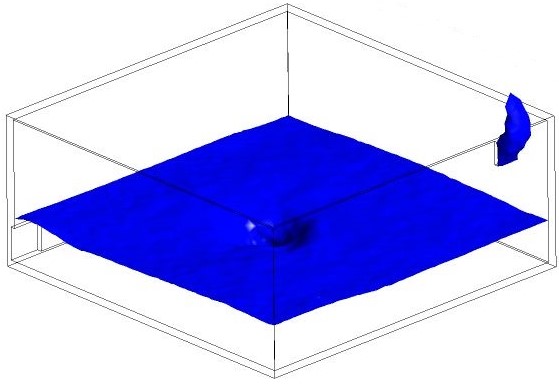}}
\subfigure[Upper layer volume defined from the iso-surface at constant temperature.]{\label{fig:isovol_w025} \includegraphics[width=0.4\textwidth]{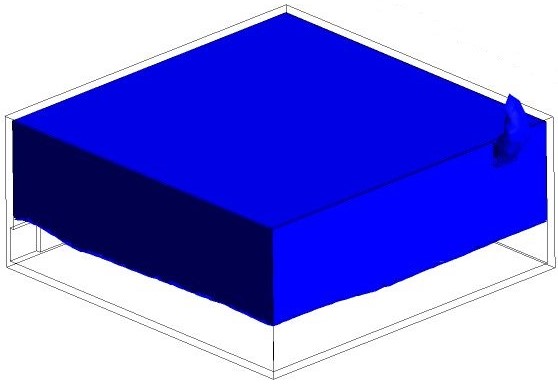}}
\caption{Iso-surface and the corresponding buoyant layer for a ideal source and a wind speed of 0.25\,m/s.}
\label{fig:isobuilding_w025}
\end{figure}

\noindent Once this temperature is know, an iso-surface at a temperature equal to $T_{int}$ is created inside the building as it is shown in Fig.~\ref{fig:isosurf_w025}. The upper (buoyant) layer is defined as the volume contained between the iso-surface and the ceiling of the building (Fig.~\ref{fig:isovol_w025}). Finally, the temperature is averaged over the upper layer.

\noindent In the case of a mixed regime, since the interface height does not exist but the air flows chaotically inside the room, the inner temperature is computed as an average over the entire room. To do so, the cells composing the inside of the building are attributed a ``+1'' value whether the external atmosphere's cells are defined at ``0.'' It is then possible to average the temperature over the cells mentioned previously.

\end{document}